\documentclass[10pt]{article}
\usepackage{pdfpages}
\usepackage{amsmath}
\usepackage{amssymb}
\usepackage{amsfonts}
\usepackage{lscape}
\usepackage{lineno}
\usepackage{setspace}
\usepackage{amssymb,amsfonts,amsmath}

\usepackage{cite}
\usepackage{sidecap}
\sidecaptionvpos{figure}{c}
\usepackage{color} 
\usepackage{booktabs}
\usepackage{stfloats}

\usepackage{blindtext}
\usepackage{centernot}
\usepackage{mathtools}
\usepackage{ stmaryrd }

\usepackage[labelfont=bf,labelsep=period,justification=justified]{caption}

\makeatletter
\renewcommand{\@biblabel}[1]{\quad#1.}
\makeatother

\date{}

\begin{document}

\title{Decomposing variability in protein levels from noisy expression, genome duplication and partitioning errors during cell-divisions}

\author{M. Soltani$^{1}$, C. A. Vargas-Garcia$^{1}$, D. Antunes$^{2}$,
A. Singh$^{1,3,4,5\ast}$}

\maketitle

\noindent {\bf{1} Electrical and Computer Engineering, University of Delaware, Newark, DE.
\\
\bf{2} Mechanical Engineering, Eindhoven University of Technology, Netherlands.
\\
\bf{3} Biomedical Engineering, University of Delaware, Newark, DE.\\
\bf{4} Mathematical Sciences, University of Delaware, Newark, DE. \\
\bf{5} Center for Bioinformatics and Computational Biology, University of Delaware, Newark, DE.
}\\

%


\noindent {$\ast$ E-mail: Corresponding absingh@udel.udu }

\section*{Abstract}

\paragraph*{Inside individual cells, expression of genes is inherently stochastic and manifests as cell-to-cell variability or noise in protein copy numbers. Since proteins half-lives can be comparable to the cell-cycle length, randomness in cell-division times generates additional intercellular variability in protein levels. Moreover, as many mRNA/protein species are expressed at low-copy numbers, errors incurred in partitioning of molecules between the mother and daughter cells are significant. We derive analytical formulas for the total noise in protein levels for a general class of cell-division time and partitioning error distributions. Using a novel hybrid approach the total noise is decomposed into components arising from i) stochastic expression; ii) partitioning errors at the time of cell-division and iii) random cell-division events. These formulas reveal that random cell-division times not only generate additional extrinsic noise but also critically affect the mean protein copy numbers and intrinsic noise components. Counter intuitively, in some parameter regimes noise in protein levels can decrease as cell-division times become more stochastic. Computations are extended to consider genome duplication, where the gene dosage is increased by two-fold at a random point in the cell-cycle. We systematically investigate how the timing of genome duplication influences different protein noise components. 
Intriguingly, results show that noise contribution from stochastic expression is minimized at an optimal genome duplication time. Our theoretical results motivate new experimental methods for decomposing protein noise levels from single-cell expression data. Characterizing the contributions of individual noise mechanisms will lead to precise estimates of gene expression parameters and techniques for altering stochasticity to change phenotype of individual cells.}

\section*{Key index words or phrases}
Single cell, stochastic gene expression, cell-division, moment dynamics, hybrid models, noise decomposition, partitioning errors
\section{Introduction}
The level of a protein can deviate considerably from cell-to-cell, in spite of the fact that cells are genetically-identical and are in the same extracellular environment \cite{bkc03,rao05,Neuert01022013}. This intercellular variation or noise in protein counts has been implicated in diverse processes such as corrupting functioning of gene networks \cite{lps07,fhg04,leh08}, driving probabilistic cell-fate decisions \cite{lod08,arm98,wbt05,wds08,siw09,dha14}, buffering cell populations from hostile changes in the environment \cite{Eldar:2010kk,vsk08,Kussell:2005tk,bmc04}, and causing clonal cells to respond differently to the same stimulus \cite{sac14,npv08,pal03}. An important source of noise driving random fluctuations in protein levels is stochastic gene expression due to the inherent probabilistic nature of biochemical processes \cite{rao08,keb05,mal13,Munsky:2009wt}. Recent experimental studies have uncovered additional noise sources that affect protein copy numbers. For example, the time take to complete cell-cycle (i.e., time between two successive cell-division events) has been observed to be stochastic across organisms \cite{wang_robust_2010,lambert_2015,tsukanov_2011,reshesl_2008,reshes_timing_2008,Roeder:2010vb,Zilman:2010ud,Hawkins:2009vw,sak13}. Given that many proteins/mRNAs are present inside cells at low-copy numbers, errors incurred in partitioning of molecules between the mother and daughter cells are significant \cite{huh_random_2011,gon13,ltr14}. Finally, the time at which a particular gene of interest is duplicated can also vary between cells \cite{zopf13,nkl15}. We investigate 
how such noise sources in the cell-cycle process combine with stochastic gene expression to generate intercellular variability in protein copy numbers (Fig.1).

\begin{figure}[h]
\centering
\includegraphics[width=1\textwidth]{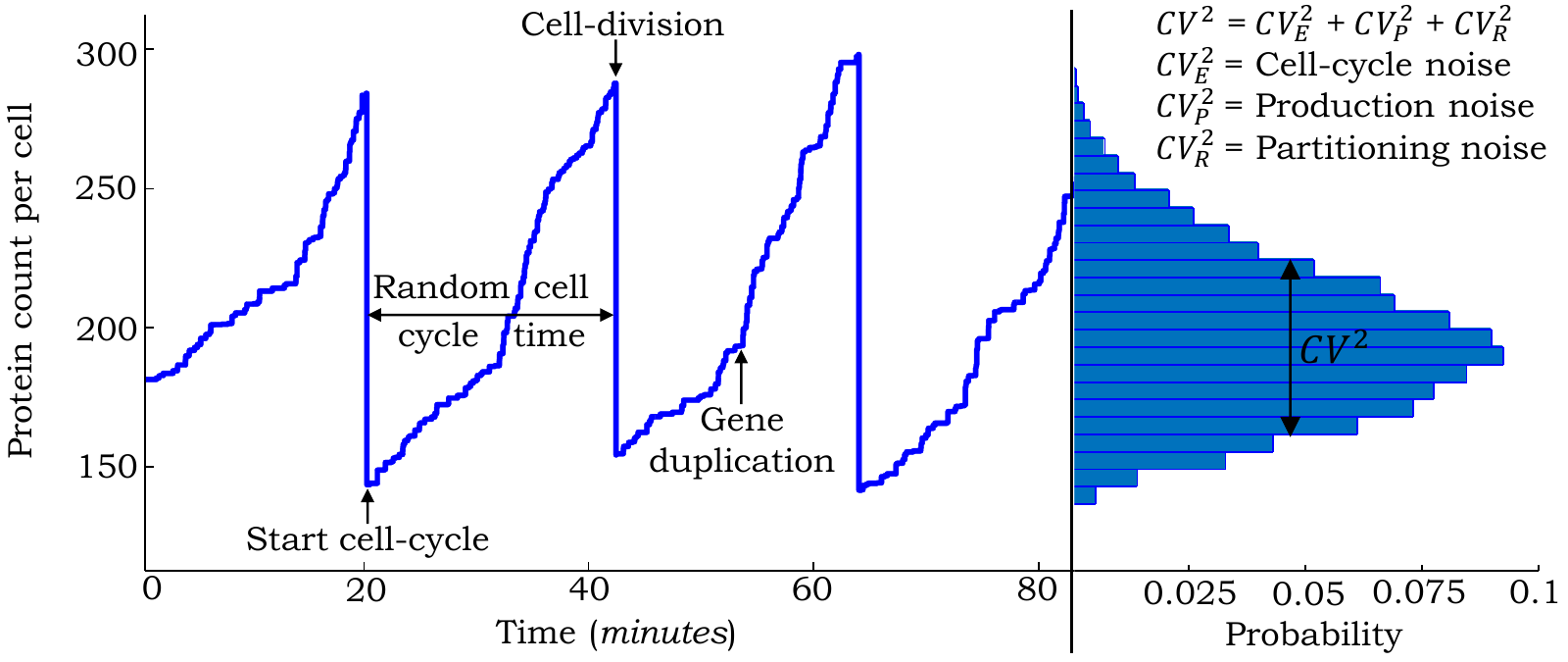}
\caption{
{\bf Sample trajectory of the protein level in a single cell with different sources of noise}. Stochastically expressed proteins accumulate within the cell at a certain rate. At a random point in the cell-cycle, gene-duplication results in an increase in production rate. Stochastic cell-division events lead to random partitioning of protein molecules between the mother and daughter 
cells with each cell receiving, on average, half the number of proteins in the mother cell just before division. The steady-state protein copy number distribution obtained from a large number of trajectories is shown on the right. The total noise in the protein level, as measured by the Coefficient of Variation ($CV$) squared can be broken into contributions from individual noise mechanisms. }
\label{fig1}
\end{figure}

Prior studies that quantify the effects of cell-division on the protein noise level have been restricted to specific cases. For example, noise computations have been done in stochastic  gene expression models, where cell-divisions occur at deterministic time intervals \cite{schwabe_2014,huh11,huh_random_2011}. Recently, we have analyzed a deterministic model of gene expression with random cell-division events \cite{antunes_2014}. Building up on this work, we formulate a mathematical model that couples stochastic expression of a stable protein with random cell-division events that follow an arbitrary probability distribution function. Moreover, at the time of cell-division, proteins are randomly partitioned between the mother and daughter cells based on a general framework that allows the partitioning errors to be higher or lower than as predicted by binomial partitioning.  For this class of models, we derive an exact analytical formula for the protein noise level as quantified by the steady-state Coefficient of Variation $(CV)$ squared. This formula is further decomposed into individual components representing contributions from different noise sources. A systematic investigation of this formula leads to novel insights, such as identification of regimes where increasing randomness in the timing of cell-division events decreases the protein noise level. 

Next, we extend the above model to include genome duplication events that increase the gene's transcription rate by two-fold (corresponding to doubling of gene dosage) 
prior to cell-division \cite{yxr06,zopf13}. To our knowledge, this is the first study integrating randomness in the genome duplication process with stochastic gene expression.
An exact formula for the protein noise level is derived for this extended model and used to investigate how the timing of duplication affects different noise components. Counter intuitively, results show that doubling of the transcription rate within the cell-cycle can lead to smaller fluctuations in protein levels as compared to a constant transcription rate through out the cell-cycle. Finally, we discuss how formulas obtained in this study can be used to infer parameters and characterize the gene expression process from single-cell studies.

\section{Coupling gene expression to cell-division}

We consider the standard model of stochastic gene expression \cite{paun05,shs08}, where mRNAs are transcribed at exponentially distributed time intervals from a constitutive gene with rate $k_x$. 
For the time being, we exclude genome duplication and the transcription rate is fixed throughout the cell-cycle.
Assuming short-lived mRNAs,
each transcription event results in a burst of proteins \cite{shs08,sih09b,jik}. The corresponding jump in protein levels is shown as
\begin{equation}
\begin{aligned}
x(t) \mapsto x(t)+ B,
\end{aligned}
\label{model0}
\end{equation}
where $x(t)$ is the protein population count in the mother cell at time $t$, $B$ is a random burst size drawn from a positively-valued distribution and represents the number of protein molecules synthesized in a single-mRNA lifetime.  Motivated by observations in \emph{E. coli} and mammalian cells, where many proteins have half-lives considerably longer than the cell-doubling time, we assume a stable protein with no active degradation \cite{alo06,Li:2010uv,sbl11}. Thus, proteins accumulate within the cell till the time of cell-division, at which point they are randomly partitioned between the mother and daughter cells. 

Let cell-division events occur at times $t_s, \ s\in \{1,2,\ldots\}$. The cell-cycle time 
\begin{align}
T  \coloneqq   t_s - t_{s-1}, \label{time}
\end{align}
follows an arbitrary positively-valued probability distribution with the following mean and coefficient of variation ($CV$) squared 
\begin{align}
&\langle T \rangle =  \langle t_s - t_{s-1} \rangle, \ \ CV^2_{T} = \frac{\langle T^2 \rangle-\langle T \rangle^2}{\langle T \rangle^2},
\end{align}
where $\langle . \rangle $ denotes expected value through out this paper.  
The random change in $x(t)$ during cell-division is given by
\begin{equation}
  x(t_s) \mapsto x_+(t_s),
\label{model}
\end{equation}
where $ x(t_s)$ and $x_+(t_s)$ denote the protein levels in the mother cell just before and after division, respectively. Conditioned on $x(t_s)$, $x_+(t_s)$ is assumed to have the following statistics
\begin{equation}
  \langle x_+(t_s)\vert  x(t_s)  \rangle= \frac{x(t_s) }{2},  \ \ \ \left \langle{x_+^2(t_s)} -\langle x_+(t_s) \rangle^2\bigg\vert  x(t_s) \right \rangle =  \frac{ \alpha x(t_s)}{4}.
  \label{division character}
\end{equation}
The first equation implies symmetric division, i.e., on average the mother cell inherits half the number protein molecules just before division. The second equation in 
\eqref{division character} describes the variance of $\langle x_+(t_s) \rangle$ and quantifies the error in partitioning of molecules through the non-negative parameter $\alpha$. For example,
$\alpha=0$ represents deterministic partitioning where $x_+(t_s) =x(t_s)/2$ with probability equal to one. A more realistic model for partitioning 
is each molecule having an equal probability of being in the mother or daughter cell \cite{ses02, berg_1978,rigney_1979}. This result in a binomial distribution
for $x_+(t_s)$
\begin{equation}
{\rm Probability}\{ x_+(t_s)=j\vert  x(t_s) \}=\frac{ x(t_s)!}{j!(x(t_s)-j)!}\left(\frac{1}{2}\right)^{x(t_s)}, \ \ j\in\{0,1,\ldots, x(t_s)\},
  \label{division character1}
\end{equation}
and corresponds to $\alpha=1$ in \eqref{division character}. Interestingly, recent studies have shown that partitioning of proteins that form clusters or multimers can
result in $\alpha>1$ in \eqref{division character}, i.e., partitioning errors are much higher than as predicted 
by the binomial distribution \cite{huh11,huh_random_2011}. 
In contrast, if molecules push each other to opposite poles of the cell, then the partitioning errors will be smaller than as predicted by \eqref{division character1} and $\alpha<1$.

The model with all the different noise mechanisms (stochastic expression; random cell-division events and partitioning errors) is illustrated in Fig. 2A and referred to as the full model.  
We also introduce two additional hybrid models \cite{sih10a,moh_14}, where protein production and partitioning are considered in their deterministic limit (Fig. 2B-C). Note that unlike the full model, where $x(t)$ takes non-negative integer values, $x(t)$ is continuous in the hybrid models. We will use these hybrid models for decomposing the protein noise level obtained from the full model into individual components representing contributions from different noise sources. However, before computing the noise, we first determine the average number of proteins as a function of the cell-cycle time distribution.

\begin{figure}[!h]
\includegraphics[width=\textwidth]{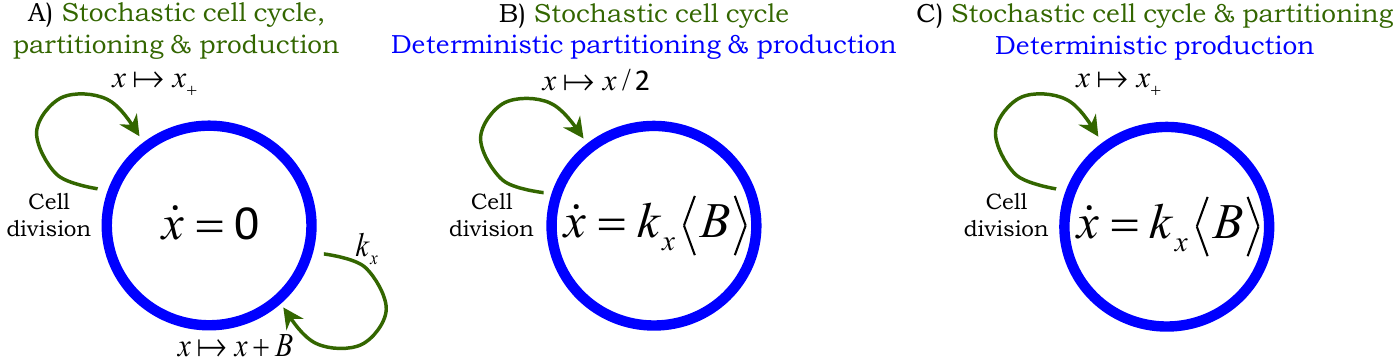} 
\caption{
{\bf  Stochastic models of gene expression with cell-division.}  Arrows denote stochastic events that change the protein level by discrete jumps as shown in  \eqref{model0} and \eqref{model}. The differential equation within the circle represents the time evolution of $x(t)$ in between events. \textbf{A)} Model with all the different sources of noise: proteins are expressed in stochastic bursts, cell-division occurs at random times, and molecules are partitioned between the mother and daughter cells based on \eqref{division character}. The trivial dynamics $\dot{x}=0$ signifies that the protein level is constant in-between stochastic events. \textbf{B)} Hybrid model where 
randomness in cell-division events is the only source of noise. Protein production is modeled deterministically through a differential equation and partitioning errors are absent, i.e., $\alpha=0$ in \eqref{division character}. \textbf{C)} Hybrid model where noise comes from both cell-division events and partitioning errors. Protein production is considered deterministically as in Fig. 2B. Since $x(t)$ is continuous here, $x_+(t_s)$ has a positively-valued continuous distribution with same mean and variance as in \eqref{division character}
}
\label{fig2}
\end{figure}

\section{Computing the average number of protein molecules}

To quantify the steady-state mean protein level we consider the full model illustrated in Fig. 2A.  It turns out that all the models shown in Fig. 2 are identical in terms of finding
$  \langle x(t)  \rangle$ and in principle any one of them could have been used. To obtain differential equations describing the time evolution of $ \langle x(t) \rangle$ we model the cell-cycle time through a phase-type distribution, which can be represented by a continuous-time Markov chain. Phase-type distributions are dense in the class of positively-valued continuous distributions, i.e., one can always construct a sequence of phase-type distributions that converges point wise to a given distribution of interest \cite{queueing01}. We use this denseness property as a practical tool for modeling the cell-cycle time. 

\subsection{Cell-cycle time as a phase-type distribution}

We consider a class of phase-type distribution that consists of a mixture of Erlang distributions. Recall that an Erlang distribution of order $i$ is the distribution of the sum of $i$ independent and identical exponential random variables. The cell-cycle time is assumed to have an Erlang distribution of order $i$ with probability $p_i, \ i=\{1,\ldots,n\}$ and  can be represented by a continuous-time Markov chain with states $G_{ij}$, $j=\{1,\ldots,i\}, \ i=\{1,\ldots,n\}$ (Fig. 3). Let Bernoulli random variables $g_{ij}=1$ if the system
resides in state $G_{ij}$ and $0$ otherwise. 
The probability of transition $G_{ij}\rightarrow G_{i(j+1)}$ in the next infinitesimal time interval $[t,t+dt)$ is given by $kg_{ij}dt$, implying that the time spent in each state $G_{ij}$ is  exponentially distributed with mean $1/k$. To summarize, at the start of cell-cycle, a state $G_{i1}, \ i=\{1,\ldots,n\}$ is chosen with probability $p_i$ and cell-division occurs after transitioning through $i$ exponentially distributed steps. Based on this formulation, the probability of a cell-division event occurring in the next time interval $[t,t+dt)$ is given by $ k p_i \sum_{j=1}^n g_{jj}dt$,
and whenever the event occurs, the protein level changes as per \eqref{model}.
Finally, the mean and the coefficient of variation squared of the cell-cycle time is obtained as 
\begin{align}
& \langle T \rangle = \sum_{i=1}^{n} p_i\frac{i}{k}, \ \ \ CV^2_{T} = \frac{1}{k}\frac{1}{\langle T \rangle } \label{phase mean cv}
\end{align}
in terms of the Markov chain parameters. Our goal is to obtain $\overline{ \langle x \rangle}:=\lim_{t \to \infty}  \langle x(t) \rangle$ as a function of $\langle T \rangle$
and $CV^2_{T}$.


\begin{SCfigure}[][h]
\includegraphics[width=0.5\textwidth]{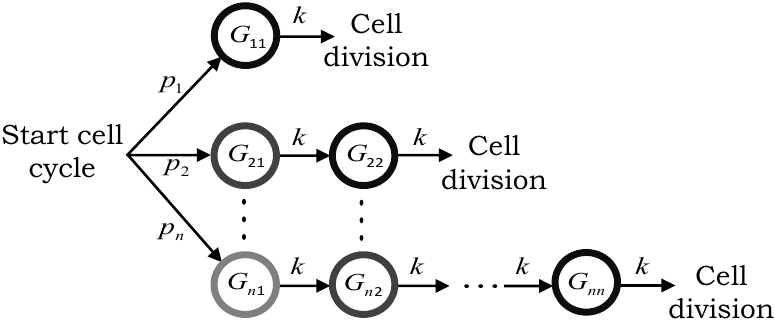} 
\caption{
{\bf A continuous-time Markov chain model for the cell-cycle time}. The cell-cycle time is assumed to follow a mixture of Erlang distributions. 
At the start of cell-cycle, a state $G_{i1}$, $i=\{1,\ldots,n\}$ is chosen with probability $p_i$. The cell-cycle transitions through states $G_{ij}, \ j=\{1,\ldots,i\}$ residing for an exponentially distributed time with mean $1/k$ in each state. Cell-division occurs after exit from $G_{ii}$ and the above process is repeated. 
}
\label{fig3}
\end{SCfigure}

\subsection{Time evolution of the mean protein level}

Time evolution of the statistical moments of $x(t)$ can be obtained from the Kolmogorov forward equations corresponding to the full model in Fig. 2A combined with the cell-division process described in Fig. 3. We refer the reader to \cite{hsi04,sih10,sih10a} for an introduction to moment dynamics for stochastic and hybrid systems. Analysis in Appendix A shows 
\begin{align}
&\frac{d\langle x \rangle}{dt}= k_x \langle B \rangle - \frac{k}{2}\left \langle \sum_{j=1}^n  x g_{jj} \right \rangle.\label{mean01}
\end{align} 
Note that the time-derivative of the mean protein level (first-order moment) is unclosed, in the sense that, it depends on the second-order moment $\langle x g_{ij} \rangle$. Typically,
approximate closure methods are used to solve moments in such cases 
\cite{gov06,lkk09,gou05,sih10,sih10a,gillespie2009,scs15}. However,
the fact that $g_{ij}$ is binary can be exploited to automatically close moment dynamics. In particular, 
since $g_{ij}\in\{0,1\}$
\begin{equation}\label{1}
\langle g_{ij}^n x^m \rangle=\langle g_{ij} x^m \rangle, \ \ \ n\in \{1,2,\ldots\}
\end{equation}
for any non-negative integer $m$. Moreover, as only a single state $g_{ij}$ can be $1$ at any time 
\begin{equation}\label{2}
\langle g_{ij} g_{rq} x^m \rangle=0, \ {\rm if} \ i\neq r \ {\rm or} \ j\neq q.
\end{equation}
Using \eqref{1} and \eqref{2}, the time evolution of $\langle x g_{ij} \rangle$ is obtained as 
\begin{subequations}\label{23}
\begin{align}
& \frac{d\langle x g_{i1} \rangle}{dt}= \frac{k_x \langle B \rangle  p_i}{i}    + \frac{k}{2}p_i\left \langle \sum_{j=1}^n x g_{jj} \right \rangle - k\langle xg_{i1} \rangle,  \label{gene01}\\
&\frac{d\langle x g_{ij} \rangle}{dt}= \frac{k_x \langle B \rangle  p_i}{i}  
- k\langle x g_{ij} \rangle + k\langle x g_{i (j-1)} \rangle,   \ j= \lbrace 2,\ldots,i \rbrace\label{gene02}
\end{align} 
\end{subequations}
and only depends on $\langle x g_{ij} \rangle$ (see Appendix A). Thus, \eqref{mean01} and \eqref{23} constitute a closed system of linear differential equations from which
moments can be computed exactly. 

To obtain an analytical formula for the average number of proteins, we start by performing a steady-state analysis of \eqref{mean01} that yields
\begin{equation}
\begin{aligned}
 \overline{ \left  \langle  \sum_{j=1}^n  x g_{jj} \right \rangle}= \frac{2 k_x\langle B \rangle}{k},\label{last gene01}
\end{aligned} 
\end{equation}
where $\overline { \langle . \rangle}$ denotes the expected value in the limit $t \to \infty$. Using \eqref{last gene01},
$\overline{\langle x g_{i1} \rangle} $ is determined from \eqref{gene01}, and then all moments $\overline{\langle x g_{ij} \rangle} $ are obtained recursively by performing 
a steady-state analysis of \eqref{gene02} for $j= \lbrace 2,\ldots,i \rbrace$. This analysis results in 
\begin{equation}
\begin{aligned}
\overline{ \langle x g_{ij} \rangle}= \frac{k_x\langle B \rangle}{k}p_i \left( 1+\frac{j}{i} \right). \label{all gene0}
\end{aligned} 
\end{equation}
Using \eqref{phase mean cv}, \eqref{all gene0} and the fact that $\sum_{i=1}^{n}\sum_{j=1}^{i}{ g_{ij}}=1$ we obtain the following expression for the mean protein level
\begin{equation}
\overline{\langle x \rangle} =\overline{\left\langle x \sum_{i=1}^{n}\sum_{j=1}^{i}{g_{ij}}\right\rangle}={\sum_{i=1}^{n}\sum_{j=1}^{i}\overline{ \langle xg_{ij} \rangle}}=\frac{k_x \langle B \rangle \langle T \rangle \left(3+CV^2_{T}\right)}{2}.\label{mean0}
\end{equation}
It is important to point that \eqref{mean0} holds irrespective of the complexity, i.e., the number of sates $G_{ij}$
used in the phase-type distribution to approximate the cell-cycle time
distribution. As expected, $\overline{\langle x \rangle}$ increases linearly with the average cell-cycle time duration $\langle T \rangle$ with longer cell-cycles resulting in more  accumulation of proteins. Consistent with previous findings, \eqref{mean0} shows that the mean protein level is also affected by the randomness in the cell-cycle times $(CV^2_{T})$ \cite{antunes_2014,wyl15}. For example, $\overline{\langle x \rangle}$ reduces by $25\%$ as $T$ changes from being exponentially distributed $(CV^2_{T}=1)$ to periodic $(CV^2_{T}=0)$
for fixed $\langle T \rangle$ fixed. Next, we determine the noise in protein copy numbers, as quantified by the coefficient of variation squared.

\section{Computing the protein noise level}

Recall that the full model introduced in Fig. 2A has three distinct noise mechanisms. Our strategy for computing the protein noise level is to first analyze the model with a single noise source, and then consider models with two and three sources. As shown below, this approach provides a systematic dissection of the protein noise level into components representing contributions
from different mechanisms. 

\subsection{Contribution from randomness in cell-cycle times}
We begin with the model shown in Fig. 2B, where noise comes from a single source - random cell-division events. For this model, the time evolution of the second-order moment of the protein copy number is obtained as 
\begin{align}
&\frac{d\langle x^2 \rangle}{dt}=
  2 k_x \langle B \rangle \langle x \rangle   - \frac{3k}{4}\left \langle \sum_{j=1}^n  x^2 g_{jj} \right \rangle ,\label{x2} 
\end{align} 
and depends on third-order moments $\langle x^2 g_{jj}  \rangle$ (see Appendix B). Using the approach introduced earlier for obtaining the mean protein level,
we close moment equations by writing the time evolution of moments $\langle x^2 g_{ij}  \rangle$. Using \eqref{1} and  \eqref{2} 
\begin{subequations}\label{x3}
\begin{align}
&\frac{d\langle x^2 g_{i1} \rangle}{dt}=  2 k_x \langle B \rangle \langle x g_{i1} \rangle  + \frac{ k}{4}p_i\left \langle \sum_{j=1}^n  x^2 g_{jj} \right \rangle  - k\langle x^2g_{i1} \rangle, \label{higher_moment_partition01}\\
&\frac{d\langle x^2 g_{ij} \rangle}{dt}=2 k_x \langle B \rangle \langle x g_{ij} \rangle 
- k\langle x^2 g_{ij} \rangle + k\langle x^2 g_{(i-1)j} \rangle, \ j=\left\lbrace 2,\ldots,i \right\rbrace.
\label{higher_moment_partition02}
\end{align} 
\end{subequations}
Note that the moment dynamics for ${ \langle x\rangle}$ and ${ \langle xg_{ij}\rangle}$ obtained in the previous section (equations \eqref{mean01} and \eqref{23}) are identical for all the models in Fig. 2, irrespective of whether the noise mechanism is modeled deterministically or stochastically. Equations \eqref{mean01}, \eqref{23}, \eqref{x2} and \eqref{x3} represent a closed set of linear differential equations and their steady-state analysis yields
\begin{align}
& \overline{ \langle x^2 g_{ij} \rangle}=\frac{ k_x^2 \langle B \rangle^2 \langle T \rangle\left(3+CV^2_{T}\right)}{3 k}p_i+\frac{2k_x^2 \langle B \rangle^2}{k}\left( \frac{j^2 +j}{i} \right)p_i.\label{all higher partition00} 
\end{align}
From \eqref{all higher partition00} 
\begin{subequations}\label{ext}
\begin{align}
&\overline{\langle x^2 \rangle} =\overline{\left\langle x^2 \sum_{i=1}^{n}\sum_{j=1}^{i}{g_{ij}}\right\rangle}={\sum_{i=1}^{n}\sum_{j=1}^{i}\overline{ \langle x^2g_{ij} \rangle}}=k_x^2 \langle B \rangle^2\frac{\langle T^3 \rangle +4  CV^2_T \langle T \rangle^3 +6\langle T \rangle^3 }{3 \langle T \rangle}, \\
&\qquad \qquad \qquad \qquad \qquad  \langle T^3 \rangle =\frac{2}{k} CV^2_{T}+ \frac{3}{k} \langle T  \rangle^2 + \langle T \rangle^3,
\end{align}
\end{subequations}
where $\langle T^3 \rangle$ is the third-order moment of the cell-cycle time. Using \eqref{ext} and the mean protein count quantified in \eqref{mean0}, we obtain the following
coefficient of variation squared 
\begin{align}
&CV^2_E= \frac{1}{27}+\frac{4\left(9\frac{\langle T^3\rangle}{\langle T \rangle^3}-9-6CV^2_
{T}-7CV^4_{T}\right)}{27\left(3+CV^2_{T}\right)^2},\label{cv_division_time0}
\end{align}
which represents the noise contribution from random cell-division events. Since cell-division is a global event that affects expression of all genes, this noise contribution 
can also be referred to as \emph{extrinsic noise} \cite{hip11,moh01,ses02,sos08,sik06}. In reality, there would be other sources of extrinsic noise, such as, fluctuations in the gene-expression machinery that we have ignored in this analysis. 

Note that $CV^2_E \to 1/27$ as $T$ approaches a delta distribution, i.e., cell divisions occur at fixed time intervals. We discuss simplifications of \eqref{cv_division_time0} in various limits. For example, if the time taken to complete cell-cycle is lognormally distributed, then
\begin{align}
\frac{\langle T^3\rangle}{\langle T \rangle^3} &= \left(1+ CV^2_{T}\right)^3 \implies CV^2_E= \frac{1}{27}+\frac{4\left(21CV^2_
{T}+20CV^4_{T}+9CV^6_{T}\right)}{27\left(3+CV^2_{T}\right)^2}
\end{align}
and extrinsic noise monotonically increases with $CV^2_{T}$. If fluctuations in $T$ around $\langle T \rangle$ are small, then using Taylor series 
\begin{align}
\langle T^3\rangle/\langle T \rangle^3 \approx 1+ 3 CV^2_{T}. \label{app}
\end{align}
Substituting \eqref{app} in \eqref{cv_division_time0} and ignoring $CV^4_{T}$ and higher order terms yields
\begin{equation}
CV^2_E\approx \frac{1}{27} +\frac{28CV^2_{T }}{81},
\end{equation}
where the first term is the extrinsic noise for ${CV^2_{T} \to 0}$ and the second term is the additional noise due to random cell-division events.

\subsection{Contribution from partitioning errors}
Next, we consider the model illustrated in Fig. 2C with both random cell-division events and partitioning of protein between the mother and daughter cells. Thus, the protein noise level here represents the contribution from both these sources. Analysis in Appendix C shows that the time evolution of $\langle x^2 \rangle$ and $\langle x^2 g_{ij} \rangle$
are given by 
\begin{subequations}\label{211}
\begin{align}
&\frac{d\langle x^2 \rangle}{dt}=
2 k_x \langle B \rangle \langle x \rangle 
+ \frac{1}{4}\alpha k\left \langle\sum_{j=1}^n x  g_{jj} \right \rangle-\frac{3}{4} k \left \langle\sum_{j=1}^n x^2  g_{jj} \right \rangle,  \label{x2 partirion0} \\
&\frac{d\langle x^2 g_{i1} \rangle}{dt}= 2 k_x \langle B \rangle \langle x g_{i1} \rangle + \frac{ k}{4} p_i\left \langle \sum_{j=1}^n  x^2 g_{jj} \right \rangle +\frac{1}{4} \alpha k p_i\left \langle \sum_{j=1}^n  x g_{jj}\right \rangle  - k\langle x^2g_{i1} \rangle,  \label{x2g1 partition0} \\
&\frac{d\langle x^2 g_{ij} \rangle}{dt}=2 k_x \langle B \rangle \langle x g_{ij} \rangle 
- k\langle x^2 g_{ij} \rangle + k\langle x^2 g_{(i-1)j} \rangle, \ j=\left\lbrace 2,\ldots,i \right\rbrace.
\label{x2g2 partition0}
 \end{align} 
\end{subequations}
Note that \eqref{x2 partirion0}-\eqref{x2g1 partition0} are slightly different from their counterparts obtained in the previous section (equations \eqref{x2} and \eqref{higher_moment_partition01})  with additional terms that depend
on $\alpha$, where $\alpha$ quantifies the degree of partitioning error as defined in \eqref{division character}. As expected, \eqref{211} reduces to \eqref{x2}-\eqref{x3} when
$\alpha=0$ (i.e., deterministic partitioning). Computing $\overline{ \langle x^2g_{ij} \rangle}$ by performing a steady-state analysis of \eqref{211} and using a similar approach as in \eqref{ext} we obtain
\begin{equation}
\overline{\langle x^2 \rangle}=\frac{\langle T^3 \rangle +4  CV^2_T \langle T \rangle^3 +6\langle T \rangle^3 }{3 \langle T \rangle}+\frac{2 \alpha k_x\langle B \rangle \langle T \rangle}{3}.
\end{equation}
Finding $CV^2$ of the protein level and subtracting the extrinsic noise found in \eqref{cv_division_time0} yields
\begin{align}
&CV^2_R =  \frac{4 \alpha }{3(3+CV^2_{T})}\frac{1}{\overline{ \langle x \rangle}},
\label{cv_partitioning0}
\end{align}
where $CV^2_R$ represents the contribution of partitioning errors to the protein noise level. Intriguingly, while $CV^2_R$ increases with $\alpha$, it decrease with 
$CV^2_{T}$. Thus,  as cell-division times become more random for a fixed  $\langle T \rangle$ and $\overline{ \langle x \rangle}$, the noise contribution from partitioning errors decrease.

\begin{figure}[!b]
\centering
\includegraphics[width=0.8\textwidth]{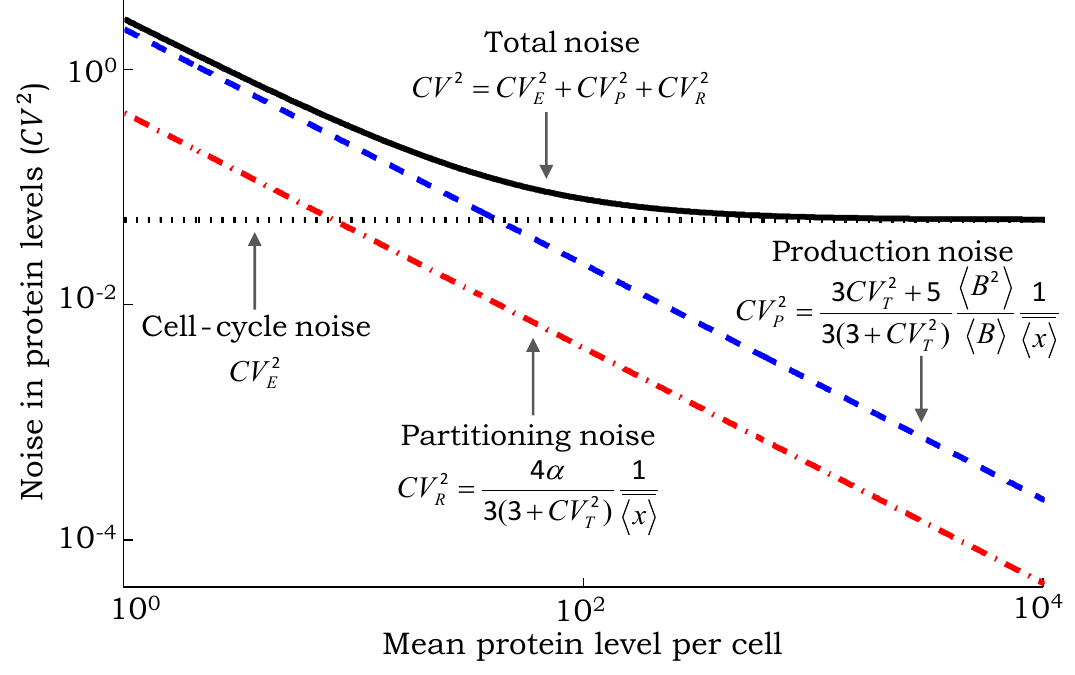} 
\caption{
{\bf Scaling of noise as a function of the mean protein level for different mechanisms}. The contribution of random cell-division events to the noise in protein copy numbers (extrinsic noise) is invariant of the mean. In contrast, contributions from partitioning errors at the time of cell-division (partitioning noise) and stochastic expression (production noise) scale inversely with the mean. The scaling factors are shown as a function of the protein random burst size $B$, noise in cell-cycle time ($CV^2_{T}$) and magnitude of partitioning errors quantified by $\alpha$ (see \eqref{division character}). With increasing mean level the total noise first decreases and then reaches a baseline that corresponds to extrinsic noise. For this plot,  $B$ is assumed to be geometrically-distributed with mean $\langle B \rangle = 1.5$, $CV^2_{T}=0$  and $\alpha=1$ (i.e., binomial partitioning). }
\label{fig5}
\end{figure}

\subsection{Contribution from stochastic expression}
Finally, we consider the full model in Fig. 2A with all the three different noise sources. For this model, moment dynamics is obtained as (see Appendix D) 
\begin{subequations} \label{213}
\begin{align}
&\frac{d\langle x^2 \rangle}{dt}=
k_x \langle B^2 \rangle +  2 k_x \langle B \rangle \langle x \rangle  +\frac{1}{4} \alpha{k}\left \langle \sum_{j=1}^n  x g_{jj}\right  \rangle  - \frac{3k}{4}\left \langle \sum_{j=1}^n  x^2 g_{jj} \right \rangle ,\label{x2_birth0} \\
&\frac{d\langle x^2 g_{i1} \rangle}{dt}=  \frac{k_x\langle B^2 \rangle p_i}{i}+2 k_x \langle B \rangle \langle x g_{i1} \rangle  + \frac{ k}{4}p_i\left \langle \sum_{j=1}^n  x^2 g_{jj} \right \rangle + \frac{1}{4} \alpha{ k}p_i\left \langle \sum_{j=1}^n  x g_{jj}\right \rangle - k\langle x^2g_{i1} \rangle, \label{x2g1_birth0}\\
&\frac{d\langle x^2 g_{ij} \rangle}{dt}=\frac{k_x\langle B^2 \rangle p_i}{i}+2 k_x \langle B \rangle \langle x g_{ij} \rangle 
- k\langle x^2 g_{ij} \rangle + k\langle x^2 g_{(i-1)j} \rangle, \ j= \left\lbrace 2,\ldots,i \right\rbrace.
\label{x2gi_birth0}
\end{align} 
\end{subequations}
Compared to \eqref{211}, \eqref{213} has additional terms of the form $k_x\langle B^2 \rangle$, where $\langle B^2 \rangle$ is the second-order moment of the protein burst size in \eqref{model0}. Performing an identical analysis as before we obtain
\begin{equation}
\overline{\langle x^2 \rangle}=\frac{\langle T^3 \rangle +4  CV^2_T \langle T \rangle^3 +6\langle T \rangle^3 }{3 \langle T \rangle} +\frac{2 \alpha k_x\langle B \rangle \langle T \rangle}{3}+\frac{ k_x \langle B^2\rangle  \langle T \rangle  (3 CV^2_T+5)}{2},
\end{equation}
which yields the following total protein noise level
\begin{equation}
\begin{aligned}
CV^2=&CV^2_E+CV^2_R + CV^2_P=CV^2_E+ \underbrace{ \overbrace{\frac{4\alpha}{3(3+CV^2_{T})}\frac{1}{\overline{ \langle x \rangle}}}^{\text{Partitioning noise }(CV^2_R)}+\overbrace{\frac{3CV^2_{T } +5}{3(3+CV^2_{T })}\frac{\langle B^2 \rangle}{\langle B\rangle}\frac{1}{\overline{ \langle x \rangle} }}^{\text{Production noise }(CV^2_P)}}_{\text{Intrinsic noise}
},
\label{total noise}  
\end{aligned} 
\end{equation}
that can be decomposed into three terms. The first is the extrinsic noise $CV^2_E$ representing the contribution from random cell-division events and given by \eqref{cv_division_time0}. The second term $CV^2_R$ is the 
contribution from partitioning errors determined in the previous section (partitioning noise), and the final term $CV^2_P$ is the additional noise representing the contribution from stochastic expression (production noise). We refer to the sum of the contributions from partitioning errors and stochastic expression as \emph{intrinsic noise}. These intrinsic and extrinsic noise components are generally obtained experimentally using the dual-color assay that measures the correlation in the expression of two identical copies of the gene \cite{ses02}.

Interestingly, for a fixed mean protein level $\overline{ \langle x \rangle}$, $CV^2_{T }$ has opposite effects on $CV^2_R$ and $CV^2_P$. While $CV^2_R$ monotonically decreases with increasing $CV^2_{T }$, $CV^2_P$ increases with $CV^2_{T }$. It turns out that in certain cases these effects can cancel each other out. For example, when $B=1$ with probability one, i.e., proteins are synthesized one at a time at exponentially distributed time intervals and $\alpha=1$ (binomial partitioning) 
\begin{equation}
CV^2= CV^2_E+ \frac{4}{3(3+CV^2_{T})}\frac{1}{\overline{ \langle x \rangle}}+ \frac{3CV^2_{T } +5}{3(3+CV^2_{T })}\frac{1}{\overline{ \langle x \rangle} }=  CV^2_E+ \frac{1}{\overline{ \langle x \rangle} }.
\label{total noise11}  
\end{equation}
In this limit the intrinsic noise is always 1/Mean irrespective of the cell-cycle time distribution $T$ \cite{huh_random_2011}. Note that the average number of proteins itself depends on $T$
as shown in \eqref{mean0}. Another important limit is $CV^2_{T} \to 0$, in which case \eqref{total noise} reduces to 
\begin{equation}
\begin{aligned}
 CV^2\approx \underbrace{ \overbrace{\frac{1}{27}}^{ CV^2_E}}_{\text{Extrinsic noise} }+  \underbrace{ \overbrace{\frac{4\alpha}{9}\frac{1}{\overline{ \langle x \rangle}}}^{CV^2_R}+\overbrace{\frac{5}{9}\frac{\langle B^2 \rangle}{\langle B\rangle}\frac{1}{\overline{ \langle x \rangle} }}^{CV^2_P}}_{\text{Intrinsic noise} },
\label{total noise1}  
\end{aligned} 
\end{equation}
and is similar to the result obtained in \cite{schwabe_2014} for deterministic cell-division times and binomial partitioning. 

Fig. 4 shows how different protein noise components change as a function of the mean protein level as the gene's transcription rate $k_x$ is modulated. The extrinsic noise is primarily determined by the distribution of the cell-cycle time and is completely independent of the mean. In contrast, both $CV^2_R$ and $CV^2_P$ scale inversely with the mean, albeit with different scaling factors (Fig. 4). This observation is particularly important since many single-cell studies in \emph{E. coli}, yeast and mammalian cells have found the protein noise levels to scale inversely with the mean across different genes \cite{otk02,ngi06,src10,bpm06}. Based on this scaling it is often assumed that the observed cell-to-cell variability in protein copy numbers is a result of stochastic expression. However, as our results show, noise generated thorough partitioning errors is also consistent with these experimental observations
and  it may be impossible to distinguish between these two noise mechanisms based on protein $CV^2$ versus mean plots unless $\alpha$ is known.

\section{Quantifying the effects of gene-duplication on protein noise}

The full model introduced in Fig. 2 assumes that the transcription rate (i.e., the protein burst arrival rate) is constant throughout the cell-cycle. This model is now extended to incorporate gene duplication during cell cycle, which is assumed to create a two-fold change in the burst arrival rate (Fig. 5). As a result of this, accumulation of proteins will be bilinear as illustrated in Fig. 1. We divide the cell-cycle time $T$ into two intervals: time from the start of cell-cycle to gene-duplication ($T_1$), and time from gene-duplication to cell-division ($T_2$). $T_1$ and $T_2$ are independent random variables that follow arbitrary distributions modeled through phase-type processes (see Fig. S2 in the Supplementary Information).  The mean cell-cycle duration and its noise can be expressed as
\begin{equation}
\begin{aligned}
&\langle {T} \rangle= \langle T_1 \rangle + \langle T_2 \rangle, \ \ \beta=\frac{\langle T_1 \rangle}{\langle {T} \rangle},
& CV^2_{{T}}=\beta^2 CV^2_{T_1}+(1-\beta)^2 CV^2_{T_2},
\end{aligned}
 \label{gene dup. time}
\end{equation}
where $CV^2_{X}$ denotes the coefficient of variation squared of the random variable $X$. An important variable in this formulation is $\beta$, which represents the average time of gene-duplication normalized by the mean cell-cycle time. Thus, $\beta$ values close to $0 \ (1)$ imply that the gene is duplicated early (late) in the cell-cycle process. Moreover, the noise in the gene-duplication time is controlled via $CV^2_{T_1}$. 
\begin{SCfigure}[][!h]
\includegraphics[width=0.5\textwidth]{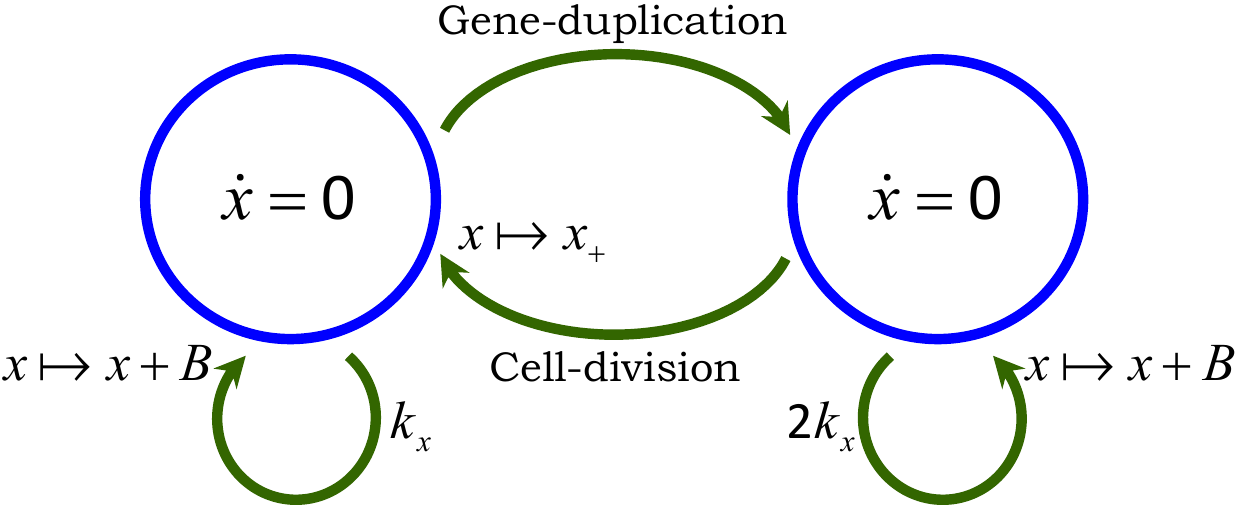} 
\caption{ \hspace{-8mm} {\bf Model illustrating stochastic expression together with random gene-duplication and cell-division events.} At the start of cell-cycle, protein production occurs in stochastic bursts with rate $k_x$. Genome duplication occurs at a random point $T_1$ within the cell-cycle and increases the burst arrival rate to $2k_x$. Cell-division occurs after time $T_2$ from genome duplication, at which point the burst arrival rate reverts back to $k_x$ and proteins are randomly partitioned between cells based on 
\eqref{model}.
}
\label{fig6}
\end{SCfigure} 

We refer the reader to Appendix E for a detailed analysis of the model in Fig. 5 and only present the main results on the protein mean and noise levels. The steady-state mean protein count is given by
\begin{equation}
\begin{aligned}
\overline{ \langle x \rangle}=\frac{k_x \langle B \rangle \langle {T_1} \rangle \left(4-\beta +  \beta CV^2_{T_1} \right)}{2}+ k_x \langle B \rangle \langle {T_2} \rangle \left(3-\beta +  (1-\beta) CV^2_{T_2} \right) ,\label{mean}
\end{aligned}
\end{equation}
and decreases with $\beta$, i.e., a gene that duplicates early has on average, more number of proteins. When $\beta=1$, then the transcription rate is $k_x$ throughout the cell-cycle and we recover the mean protein level obtained in \eqref{mean0}. Similarly, when $\beta=0$ the transcription rate is $2k_x$ and we obtain twice the amount as in \eqref{mean0}. As per our earlier observation, more randomness in the timing of genome duplication and cell-division (i.e., higher $CV^2_{T_1}$ and $CV^2_{T_2}$ values) increases $\overline{ \langle x \rangle}$.

Our analysis shows that the total protein noise level can be decomposed into three components 
\begin{equation}\label{gened}
CV^2=CV^2_E+CV^2_R+ CV^2_P
 \end{equation}
 where $CV^2_E$ is the extrinsic noise from random genome-duplication and cell-division events. Given its complexity, we refer the reader to equation (100) in Appendix E2 for an exact formula for $CV^2_E$. Moreover, the intrinsic noise, which represents the sum of contributions from partitioning errors ($CV^2_R$) and stochastic expression ($CV^2_P$) is obtained as 
 \begin{small}
 \begin{equation}
\begin{aligned}
&CV^2_R+CV^2_P=\\
&\overbrace{\frac{4\alpha (2-\beta)}{3\left((\beta^2 -4 \beta +6)  + \beta^2CV^2_{T_1}+ 2(1-\beta)^2 CV^2_{T_2} \right)}\frac{1}{\overline{ \langle x \rangle}}}^{CV^2_R}+\overbrace{  \frac{(10-8\beta+3\beta^2) + 6(1-\beta)^2 CV^2_{T_2} + 3 \beta^2 CV^2_{T_1} }{3\left((\beta^2 -4 \beta +6)  +\beta^2CV^2_{T_1} + 2 (1-\beta)^2 CV^2_{T_2} \right)}\frac{\langle B^2 \rangle}{\langle B\rangle}\frac{1}{\overline{ \langle x \rangle}}}^{CV^2_P}.
\label{cv_partitioning1}
\end{aligned}
\end{equation}
\end{small}
Note that for $\beta=0$ and $1$, we recover the intrinsic noise level in \eqref{total noise} from \eqref{cv_partitioning1}. Interestingly, for
$B=1$ with probability $1$ and $\alpha=1$, the intrinsic noise is always $1/\text{Mean}$ irrespective of the values chosen for $CV^2_{T_1}$, $CV^2_{T_2}$ and $\beta$.
For high precision in the timing of cell-cycle events ($CV_{T_1} \to 0, CV_{T_2} \to 0$)
\begin{align}\label{111}
CV^2 \approx
\underbrace{ \overbrace{ \frac{4-3 (\beta-2)^2 \beta^2}{3\left(\beta^2 -4 \beta +6 \right)^2}}^{CV^2_E}
}_{\text{Extrinsic noise}  }&+   \underbrace{\overbrace{\frac{4\alpha (2-\beta)}{3\left(\beta^2 -4 \beta +6 \right)}\frac{1}{\overline{ \langle x \rangle}}}^{CV^2_R} + \overbrace{ \frac{(10-8\beta+3\beta^2)}{3\left(\beta^2 -4 \beta +6 \right)}\frac{\langle B^2 \rangle}{\langle B\rangle}\frac{1}{\overline{ \langle x \rangle}}}^{CV^2_P}}_{\text{Intrinsic noise}}, 
\end{align}
where mean protein  level is given by
\begin{align}
 \overline{ \langle x \rangle} \approx &\frac{k_x \langle B \rangle \langle {T_1} \rangle \left(4-\beta \right)}{2}+ k_x \langle B \rangle \langle {T_2} \rangle \left(3-\beta  \right).
\end{align}
We investigate how different noise components in \eqref{111} vary with $\beta$ as the mean protein level is held fixed by changing $k_x$. Fig. 6 shows that $CV^2_P$ follows a U-shaped profile with the optima occurring at $\beta=2-\sqrt{2} \approx 0.6$ and the corresponding minimum value being $\approx 5\%$ lower that its value at $\beta=0$. An implication of this result is that if stochastic expression is the dominant noise source, then gene-duplication can result in slightly lower protein noise levels. In contrast to $CV^2_P$, $CV^2_R$ has a maxima
at $\beta=2-\sqrt{2} $ which is $\approx 6\%$ higher than its value at $\beta=0$ (Fig. 6). Analysis in Appendix E5 reveals that $CV^2_R$ and $CV^2_R$ follow the same qualitative shapes as in Fig. 6  for non-zero $CV^2_{T_1}$ and  $CV^2_{T_2}$. Interestingly, when 
$CV^2_{T_1}=CV^2_{T_2}$, the maximum and minimum values of $CV^2_R$ and $CV^2_P$ always occur at $\beta=2-\sqrt{2}$ albeit with different optimal values than Fig. 6 (see Fig. S3 in the Supplementary Information). For example, if $CV^2_{T_1}=CV^2_{T_2}=1$ (i.e., exponentially distributed $T_1$ and $T_2$), then the maximum value of $CV^2_R$ is $20\%$ higher and the minimum value of $CV^2_P$ is $10\%$ lower than their respective value for $\beta=0$. Given that the effect of changing $\beta$ on $CV^2_P$ and $CV^2_R$ is small and antagonistic, the overall affect of genome duplication on intrinsic noise may be minimal and hard to detect experimentally. 

\begin{figure}[h]
\centering
\includegraphics[width=0.75\textwidth ,clip]{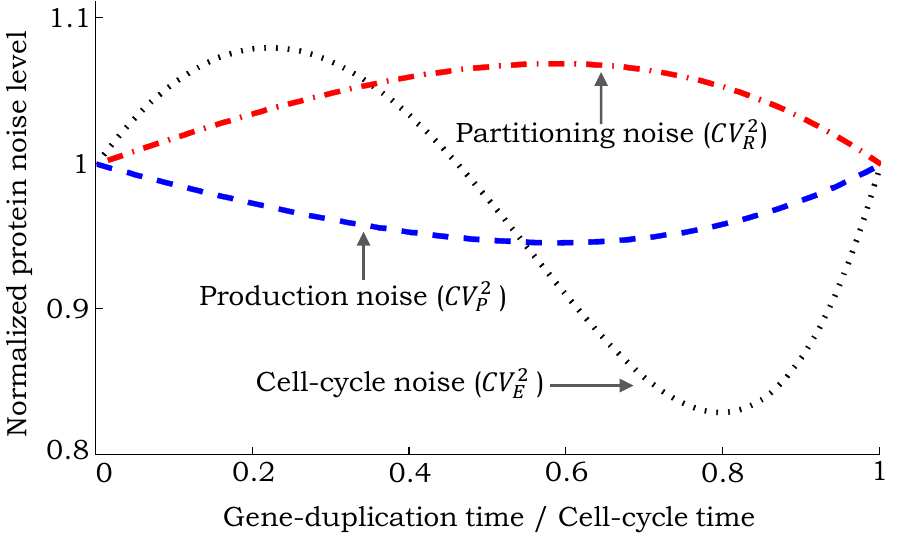} 
\caption{
{\bf Contributions from different noise sources as a function of the timing of genome duplication for $CV^2_{T_1}=CV^2_{T_2}=0$}. Different noise components in \eqref{111} are plotted as a function 
of $\beta$, which represents the fraction of time within the cell-cycle at which gene-duplication occurs. The mean protein level is held constant by simultaneously changing the transcription rate $k_x$. Noise levels are normalized by their respective value at $\beta=0$.
The noise contribution from partitioning errors is maximized at $\beta \approx 0.6$. In contrast, the contribution from stochastic expression is minimum at $\beta \approx 0.6$. The extrinsic noise contribution from random gene-duplication and cell-division events is maximum at $\beta \approx 0.2$ and minimum at $\beta \approx 0.8$.}
\label{fig7}
\end{figure}

\section{Discussion}
We have investigated a model of protein expression in bursts coupled to discrete gene-duplication and cell-division events. The novelty of our modeling framework lies in describing the size of protein bursts, $T_1$ (time between cell birth and gene duplication), $T_2$ (time between gene duplication and cell division) and partitioning of molecules during cell division through arbitrary distributions. Exact formulas connecting the protein mean and noise levels to these underlying distributions were derived. Furthermore, the protein noise level, as measured by the coefficient of variation squared, was decomposed into three components representing contributions from gene-duplication/cell-division events, stochastic expression and random partitioning. While the first component is independent of the mean protein level, the other two components are inversely proportional to it. Key insights obtained are as follows:
\begin{itemize}
\item The mean protein level is affected by  both the first and second-order moments of $T_1$ and $T_2$. In particular, randomness in these times (for a fixed mean) increases the average protein count. 
\item Random gene-duplication/cell-division events create an extrinsic noise term which is completely determined by moments of $T_1$ and $T_2$ up to order three.
\item The noise contribution from partitioning errors decreases with increasing randomness in $T_1$ and $T_2$. Thus, if $\overline{\langle x \rangle}$ is sufficiently small and $\alpha$ is large compared to $B$ in \eqref{cv_partitioning1}, increasing noise in the timing of cell-cycle events decreases the total noise level. 
\item Genome duplication has counter intuitive effects on the protein noise level (Fig. 6). For example, if stochastic expression is the dominant source of noise, then doubling of transcription due to duplication results in lower noise as compared to constant transcription throughout the cell-cycle. 
\item For a non-bursty protein production process ($B=1$) and binomial partitioning ($\alpha=1$), the net noise from stochastic expression and partitioning is always $1/\overline{\langle x \rangle}$, the noise level predicted by a Poisson distribution.
\end{itemize}
We discuss our results on gene duplication in further detail and how noise formulas  derived here can be used for estimating model parameters from single-cell expression data.

\subsection{Affect of gene duplication on noise level}


In this first-of-its-kind study, we have investigated how discrete  two-fold changes in the transcription rate  due do gene duplication affect the intercellular variability in protein levels. Not surprisingly, the timing of genome duplication has a strong effect on the mean protein level -- $\overline{\langle x \rangle}$ changes by two-fold depending on whether the gene duplicates early $(\beta=0)$ or late $(\beta=1)$ in the cell-cycle. In contrast, the effect of $\beta$ on noise is quite small. As $\beta$ is varied keeping $\overline{\langle x \rangle}$ fixed, noise components deviate by $\approx 10\%$ from their values at $\beta=0$ (Fig. 6). Recall that these results are for a stable protein, whose intracellular copy number accumulate
in a bilinear fashion. A natural question to ask is how would these results change for an unstable protein?

Consider an unstable protein with half-life considerably shorter than the cell-cycle duration. This rapid turnover ensures that the protein level equilibrates instantaneously after cell-division and gene-duplication events. Let $\gamma_x$ denote the protein decay rate. Then, the mean protein level before and after genome duplication is $\overline{\langle x \rangle}=k_x\langle B\rangle/\gamma_x$ and $\overline{\langle x \rangle}=2k_x\langle B\rangle/\gamma_x$, respectively. Note that in the limit of large $\gamma_x$ there is no noise contribution form partitioning errors since errors incurred at the time of cell division would be instantaneously corrected. The extrinsic noise, which can be interpreted as the protein noise level for deterministic protein production and decay is obtained as (see Appendix F)
\begin{equation}
CV^2_E= \frac{(1-\beta)\beta}{(2-\beta)^2}. \label{extrinsic}
\end{equation} 
When $\beta=0$ or $1$, the transcription rate and the protein level are constant within the cell cycle and $CV^2_E=0$. Moreover, $CV^2_E$ is maximized
at $\beta=2/3$ with a value of $1/12$. Thus, in contrast to a stable protein, extrinsic noise in an unstable protein is strongly dependent on the timing of gene duplication. Next, consider the intrinsic noise component. Analysis in Appendix F shows that the noise contribution from random protein production and decay is
\begin{equation}
CV^2_P=\frac{1}{2}\left( \frac{\langle B^2 \rangle}{\langle B\rangle}+1\right)\frac{1}{\overline{ \langle x \rangle}}, \ \  \overline{ \langle x \rangle}= \frac{k_x \langle B \rangle (2-\beta)}{\gamma_x}.
\label{production}
\end{equation} 
While the mean protein level is strongly dependent on $\beta$, the intrinsic noise Fano factor $=CV^2_P \times \langle x \rangle$ is independent of it. Thus, similar to what was observed for a stable protein, the intrinsic noise in an unstable protein is invariant of $\beta$ for a fixed $\overline{ \langle x \rangle}$. Overall, these results suggest that studies quantifying intrinsic noise in gene expression models, or using intrinsic noise to estimate model parameters (see below) can ignore the effects of gene duplication. Finally, note that the mean and noise levels obtained for an unstable protein are independent of the cell-cycle time $T$.

\subsection{Parameter inference from single-cell data}

Simple models of bursty expression and decay predict the distribution of protein levels to be negative binomial (or gamma distributed in the continuous framework) \cite{friedman2006lsd,pau05}. These distributions are characterized by two parameter -- the burst arrival rate $k_x$ and the average burst size $\langle B\rangle$, which can be estimated from measured protein mean and noise levels. This method has been used for estimating $k_x$  and $\langle B\rangle$ across different genes in \emph{E. coli} \cite{Li:2010uv,shc14}. Our detailed model that takes into account partitioning errors predicts (ignoring gene duplication effects)
\begin{equation}
\begin{aligned}
{\rm Intrinsic \ noise} =&{{\frac{4\alpha}{3(3+CV^2_{T})}\frac{1}{\overline{ \langle x \rangle}}}+{\frac{3CV^2_{T } +5}{3(3+CV^2_{T })}\frac{\langle B^2 \rangle}{\langle B\rangle}\frac{1}{\overline{ \langle x \rangle} }}}.
\label{total noise11}  
\end{aligned} 
\end{equation}
Using $CV^2_{T} \ll 1$ and a geometrically distributed $B$ \cite{gpz05,Yu17032006,cfx06,berg_1978}, \eqref{total noise11} reduces to 
\begin{equation}
\begin{aligned}
{\rm Intrinsic \ noise} =&{{\frac{4\alpha}{9}\frac{1}{\overline{ \langle x \rangle}}}+{\frac{5}{9}\frac{1+2\langle B\rangle}{\overline{ \langle x \rangle} }}}.
\label{total noise1111}  
\end{aligned} 
\end{equation}
Given measurements of intrinsic noise and the mean protein level, $\langle B\rangle$ can be estimated from \eqref{total noise1111} assuming $\alpha=1$ (i.e., binomial partitioning). Once $\langle B\rangle$
is known, $k_x$ is obtained from the mean protein level given by \eqref{mean0}. Since for many genes $\langle B\rangle\approx 0.5-5$ \cite{Li:2010uv}, the contribution of the first term
in \eqref{total noise1111} is significant, and ignoring it could lead to overestimation of $\langle B\rangle$. Overestimation would be even more severe if $\alpha$ happen to be much higher than $1$, as would be the case for proteins that form aggregates or multimers \cite{huh_random_2011}. One approach to estimate both $\langle B\rangle$ and $\alpha$ is to 
measure intrinsic noise changes in response to perturbing $\langle B\rangle$ by, for example, changing the mRNA translation rate through mutations in the ribosomal-binding sites (RBS). Consider a hypothetical scenario where the Fano Factor (intrinsic noise times the mean level) is $6$. Let mutations in the 
RBS reduces $\overline{ \langle x \rangle}$ by $50\%$, implying a $50\%$ reduction in $\langle B\rangle$. If the Fano factor changes from $6$ to $4$ due to this mutation, then $\langle B\rangle = 3.6$ and $\langle \alpha \rangle=3.25$.

Our recent work has shown that higher-order statistics of protein levels (i.e., skewness and kurtosis) or transient changes in protein noise levels in response to blocking transcription  provide additional information for discriminating between noise mechanisms \cite{ksk15,singh_transient_2014}. Up till now these studies have  ignored noise sources in the cell-cycle process. It remains to be seen if such methods can be used for separating the noise contributions of partitioning errors and stochastic expression to reliably estimate $\langle B\rangle$ and $\alpha$.

\subsection{Integrating cell size and promoter switching}

An important limitation of our modeling approach is that it does not take into account the size of growing cells. Recent experimental studies have provided important insights into the regulatory mechanisms controlling cell size  \cite{onl14,rka14,kgo13}. More specifically, studies in \emph{E. coli} and yeast argue for an ``adder" model, where cell-cycle timing is controlled so as to add a constant volume between cell birth and division \cite{am14,tbs15,csk14}. Assuming exponential growth, this implies that 
the time taken to complete cell-cycle is negatively correlated with cell size at birth. In addition, cell size also affects gene expression -- in mammalian cells transcription rates linearly increase with the cell size \cite{padovan_2015}. Thus, as cells become bigger they also produce more mRNAs to ensure gene product concentrations remains more or less constant. An important direction of future work would to explicitly include cell size with size-dependent expression and timing of cell division determined by the adder model. This formulation will for the first time, allow simultaneous investigation of stochasticity in cell size, protein molecular count and concentration. 

Our study ignores genetic promoter switching between active and inactive states, which has been shown to be a major source of noise in the expression of genes across organisms \cite{Suter:2011fk,bmf_13,rpt06,hbr12,srd12,drs12,coc14,Bothma03072014,Chubb20061018,ccg14}.  Taking into account promote switching is particularly important for genome duplication studies, where doubling the number of gene copies could lead to more efficient averaging of promoter fluctuations. Another direction of future work will be to 
incorporate this addition noise source into the modeling framework and investigate its contribution as a function of gene-duplication timing.

\appendix

\section*{Appendix}

\section{Mean of protein in the presence of cell-cycle variations}
Based on standard stochastic formulation of chemical kinetics \cite{mcq67,gil01}, the model introduced in Figure 2A coupled with phase-type distribution introduced in Figure 3 contains the following stochastic events 
\begin{figure}[h]
\centering
\includegraphics[width=0.8\textwidth]{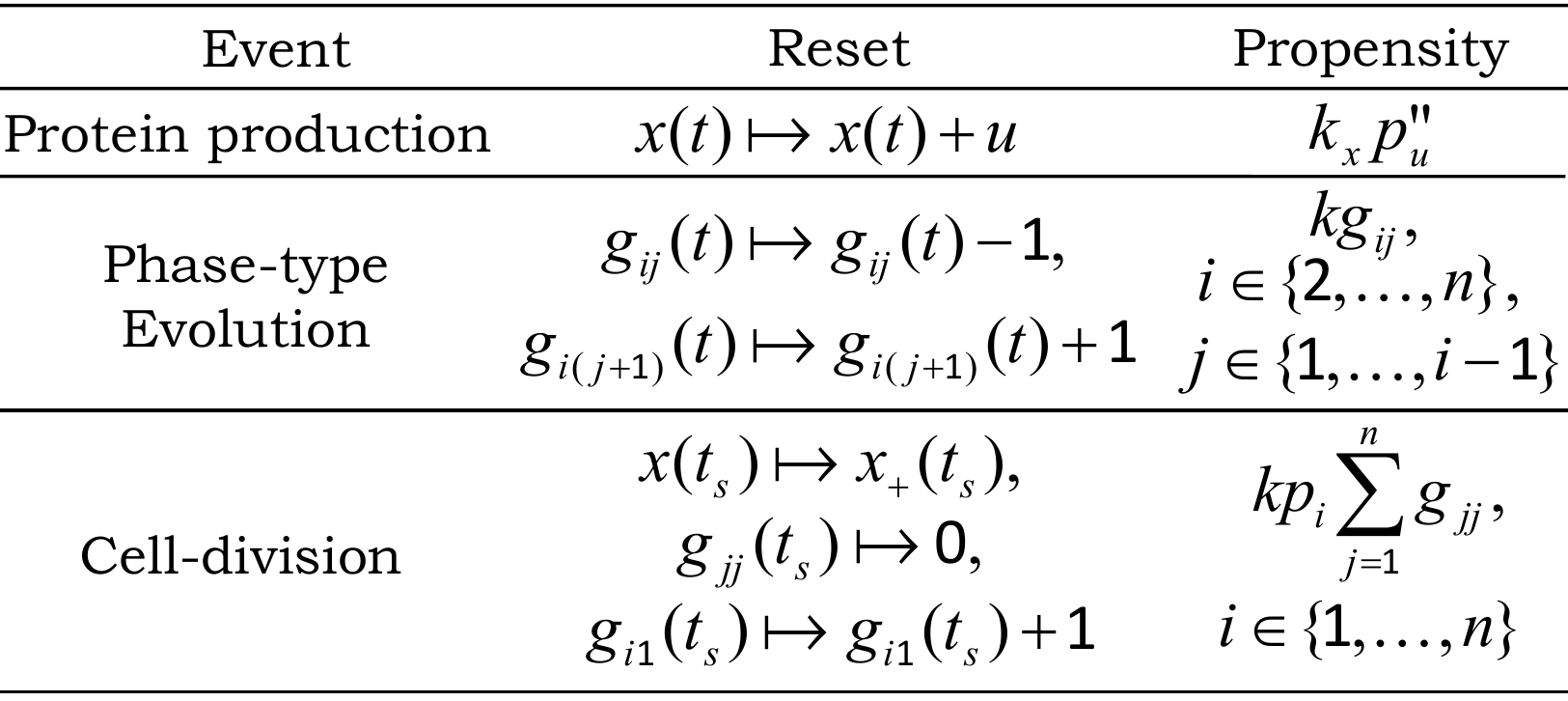} 
\end{figure}\\
\noindent Note that $x_+(t_s)$ is protein level after division, characteristics of $x_+(t_s)$ is related to protein level before division as shown in equation (5) of the main text.
Whenever an event occurs, protein level and states of phase-type distribution change based on the stoichiometries shown in the second column of the table. The third column of table shows event propensity function $f(x,g_{ij})$, which determines how often reactions occur, i.e., the probability that an event occurs in the next infinitesimal time interval $(t,t+dt]$ is $f(x,g_{ij})dt$. 
Protein production is a stochastic event which happens in bursts, each burst generates $B$ molecules where $B$ is a general random variable with distribution
\begin{equation}
{\rm Probability}\{ B=u \}= p''_u, \ \  \ u\in \{0,1, \ldots, \infty \}.
\end{equation}
The probability of having a burst in the time interval $(t,t+dt]$ is $k _x p''_u dt$. Events related to time evolution of phase-type distribution happen with a constant rate $k$. Cell-division changes both the level of protein and states of phase-type. This event contains start of new cell-cycle, hence whenever this event occurs, the last state of phase-type distribution resets to zero, and a new cell-cycle which is sum of $i$ exponentials starts with probability $p_i$; protein count level also resets to $x_+(t_s)$. The probability of cell-division and starting a new cell-cycle from state $g_{i1}$ in the time interval  $(t,t+dt]$ is $kp_i \sum_{j=1}^n dt$.

Theorem 1 of \cite{hsi04} gives the time derivative of the expected value of any function $\varphi (x,g_{ij})$ as 
\begin{equation}
\frac{d\langle \varphi (x,g_{ij}) \rangle}{dt}= \left \langle \sum_{Events}  \Delta \varphi (x,g_{ij}) \times f(x,g_{ij}) \right \rangle, 
\label{general dynamic}  
\end{equation}
where $\Delta \varphi(x,g_{ij})$ is a change in $\varphi$ when an event occurs. 
Based on this setup, mean dynamics of protein can be written by choosing $\varphi$ to be $x$
\begin{equation}
\begin{aligned}
&\frac{d\langle x \rangle}{dt}= k_x \langle B \rangle + k \left \langle \sum_{j=1}^n  ( \frac{x}{2}  -x) g_{jj} \right \rangle\Rightarrow\\
&\frac{d\langle x \rangle}{dt}= k_x \langle B \rangle - \frac{k}{2} \left \langle \sum_{j=1}^n  x g_{jj} \right \rangle,
\end{aligned} \label{x}
\end{equation}
where we replaced conditional expected value of $x_+$ by $x/2$ based on relation between statistical properties of $x_+$ and $x$ shown in equation (5).
 
Dynamics of $\langle x \rangle$ is not closed and depends to moments $\langle xg_{jj} \rangle $, hence in order to have a closed set of equations we add new moments dynamics by selecting $\varphi$ to be $ xg_{ij} $. We do it in two steps: first we write the moment dynamics of $\langle xg_{11} \rangle $
\begin{equation}
\begin{aligned}
&\frac{d\langle x g_{11} \rangle}{dt}= {k_x \langle B \rangle  \langle g_{11} \rangle }    + \frac{k}{2} p_1 \left \langle  x g_{11}^2  \right\rangle  - k p_1 \left \langle  x g_{11}^2  \right\rangle  - k  \sum_{i=2}^{n} p_i \left \langle   x g_{11} \right\rangle    . \label{gi111}
\end{aligned}
\end{equation}
In the equation (9) of the main text it has been shown that 
\begin{equation}
\begin{aligned}
\langle g_{ij}^n x^m \rangle=\langle g_{ij} x^m \rangle, \ \ \ n\in \{1,2,\ldots\},
\end{aligned}
\end{equation}
thus the term $\left \langle x g_{11}^2  \right\rangle$ will simplify as 
\begin{equation}
\left \langle xg_{11}^2  \right\rangle = \left \langle x g_{11} \right\rangle,
\end{equation}
and the dynamics of $\langle xg_{11} \rangle$ can be written as
\begin{equation}
\begin{aligned}
&\frac{d\langle x g_{11} \rangle}{dt}= k_x \langle B \rangle  \langle g_{11} \rangle    + \frac{k}{2} p_1 \left \langle   x g_{11} \right\rangle  - k  \left \langle   x g_{11} \right\rangle. \label{gi111}
\end{aligned}
\end{equation}
In the second step we write dynamics of the moments of the form $\langle xg_{ij} \rangle$ other than $\langle xg_{11} \rangle $
\begin{subequations}
\begin{align}
& \frac{d\langle x g_{i1} \rangle}{dt}= k_x \langle B \rangle  \langle g_{i1} \rangle    + k p_i\left \langle \sum_{j=1}^n (\frac{x}{2} + \frac{x}{2} g_{i1} - xg_{i1} ) g_{jj} \right\rangle - k\langle xg_{i1} \rangle ,\\
&\frac{d\langle x g_{ij} \rangle}{dt}= k_x \langle B \rangle \langle g_{ij}  \rangle
- k\langle x g_{ij} \rangle + k\langle x g_{i (j-1)} \rangle , \ \ j\in \lbrace 2,\ldots,i \rbrace,\label{gij tot}
\end{align} 
\end{subequations}
where dynamics of $\langle x g_{i1} \rangle$ can be written as 
\begin{equation}
 \frac{d\langle x g_{i1} \rangle}{dt}= \frac{k_x \langle B \rangle  p_i}{i}    + k p_i\left \langle \sum_{j=1}^n \frac{x}{2}  g_{jj} \right\rangle   + k p_i\left \langle \sum_{j=1}^n -\frac{x}{2} g_{i1} g_{jj} \right\rangle  - k\langle xg_{i1} \rangle. \label{gi11}
\end{equation}
The equation (10) in the main text shows that  
\begin{equation}
\langle g_{ij} g_{rq} x^m \rangle=0, \ {\rm if} \ i\neq r \ {\rm or} \ j\neq q,
\end{equation}
hence $\left \langle \sum_{j=1}^n \frac{x}{2} g_{i1} g_{jj} \right\rangle =0$, and equation \eqref{gi11} simplifies to
\begin{equation}
 \frac{d\langle x g_{i1} \rangle}{dt}= {k_x \langle B \rangle \langle g_{i1} \rangle}    + \frac{k}{2} p_i\left \langle \sum_{j=1}^n x  g_{jj} \right\rangle   - k\langle xg_{i1} \rangle. \label{gi1 tot}
\end{equation}

Further based on Figure 3 in the main text the probability of selecting a branch of $i$ exponentials is $p_i$, and because all the transitions happen with a constant rate $k$, hence mean of each of these $i$ states is 
\begin{equation}
\langle g_{ij} \rangle = \frac{p_i}{i}.
\end{equation}
Thus equations \eqref{gi111}, \eqref{gij tot}, and \eqref{gi1 tot} can be compactly written as shown in equation (11).

\section{Moment dynamics of hybrid model introduced in Figure 2B}
Stochastic hybrid system introduced in Figure 2B coupled with phase-type distribution contains the following stochastic events 
\begin{figure}[h]
\centering
\includegraphics[width=0.8\textwidth]{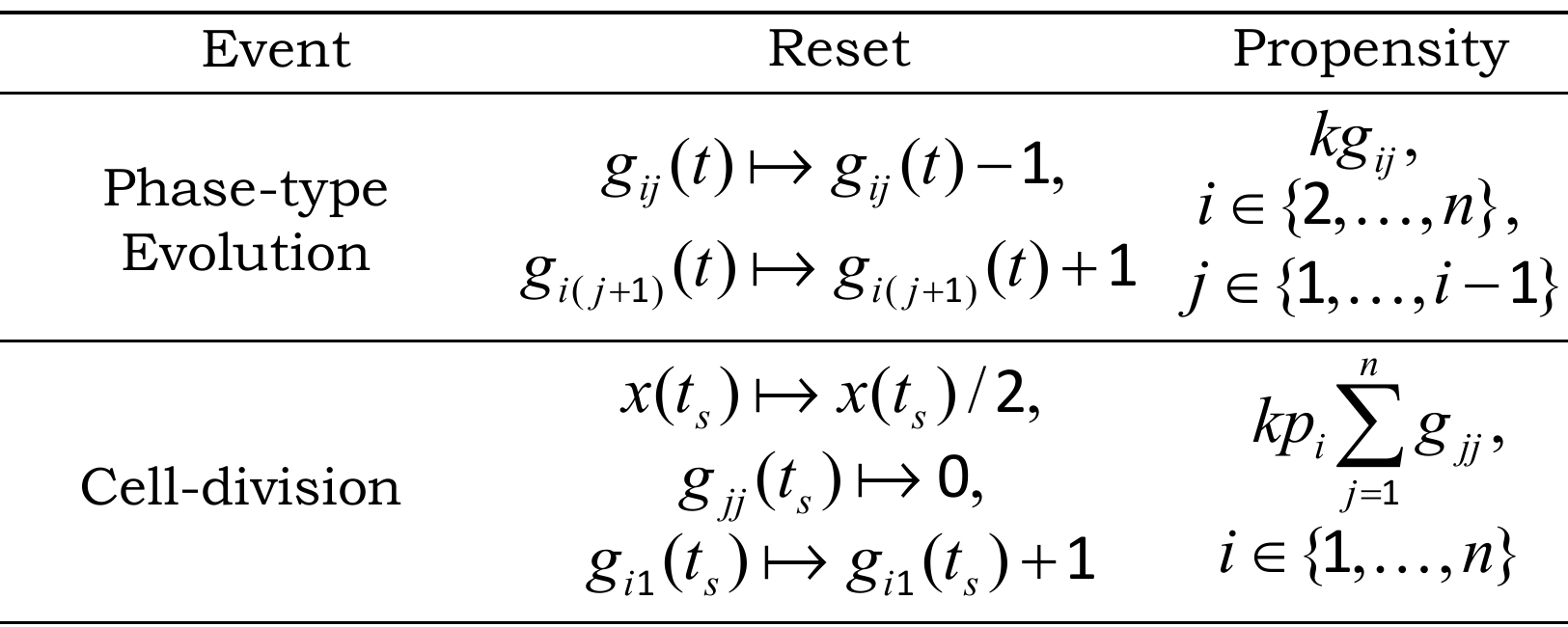} 
\end{figure}\\
and deterministic protein production dynamics
\begin{equation}
\dot{x}= k_x \langle B \rangle.
\end{equation}
Time derivative of the expected value of any function $\varphi(x,g_{ij})$ for this hybrid system can be written as \cite{hsi04}
\begin{equation}
\begin{aligned}
\frac{d\langle \varphi(x,g_{ij}) \rangle}{dt}=& \left \langle \sum_{Events}  \Delta \varphi(x,g_{ij}) \times f(x,g_{ij})\right \rangle + \left \langle \frac{\partial \varphi(x,g_{ij})}{\partial x} k_x \langle B \rangle \right \rangle,   
\label{dynnf}
\end{aligned}
\end{equation}
where the first term in the right-hand side is contributed from stochastic events and the second term is contributed from deterministic protein production dynamics. Based on this equation, the mean dynamics of the protein is calculated by choosing $\varphi$ to be $x$
\begin{equation}
\frac{d\langle x \rangle}{dt}= k_x \langle B \rangle - \frac{k}{2} \left \langle \sum_{j=1}^n  x g_{jj} \right \rangle,
\label{partitioning equations0}
\end{equation}
which is the same as equation \eqref{x}. In addition to mean, dynamics of $\langle xg_{ij} \rangle $ are also equal to their equation in the previous section.

The second order moment dynamics of protein can be expressed by choosing $\varphi$ to be $x^2$
\begin{equation}
\frac{d\langle x^2 \rangle}{dt}=
 2 k_x \langle B \rangle \langle x \rangle   +k \left \langle \sum_{j=1}^n  \left(\left( \frac{x}{2} \right) ^2-x^2 \right) g_{jj} \right \rangle ,  
\label{dynnf0}
\end{equation}
which can be simplified as
\begin{equation}
\frac{d\langle x^2 \rangle}{dt}=
 2 k_x \langle B \rangle \langle x \rangle   - \frac{3k}{4}\left \langle \sum_{j=1}^n  x^2 g_{jj} \right \rangle.
\label{partitioning equations0}
\end{equation}
In order to have a closed set of equations we select $\varphi$ to be of the form $x^2g_{ij}$. At the first step we write moment dynamics of $\langle x^2g_{11} \rangle $
\begin{equation}
\frac{d\langle x^2 g_{11} \rangle}{dt}=  2 k_x \langle B \rangle \langle x g_{11} \rangle  + \frac{k}{4} p_1 \left \langle  x^2 g_{11}^2  \right \rangle - k p_1 \left \langle  x^2 g_{11}^2  \right \rangle - k \sum_{i=2}^{n} p_i\langle x^2g_{11} \rangle. \label{x2g1}\\
\end{equation}
Based on equation (9) of the main text, the term $\left \langle x^2 g_{11}^2  \right\rangle$ simplifies as 
\begin{equation}
\left \langle x^2g_{11}^2  \right\rangle = \left \langle x^2 g_{11} \right\rangle,
\end{equation}
hence dynamics of $\langle x^2g_{11} \rangle$ will be
\begin{equation}
\begin{aligned}
\frac{d\langle x^2 g_{11} \rangle}{dt}=  2 k_x \langle B \rangle \langle x g_{11} \rangle  + \frac{k}{4} p_1 \left \langle  x^2 g_{11} \right \rangle  - k \langle x^2g_{11} \rangle. \label{x2g11}
\end{aligned}
\end{equation}
In the second step, we write dynamics of moments $\langle x^2 g_{ij} \rangle$ when $g_{ij}\neq g_{11}$
\begin{subequations}
\begin{align}
&\frac{d\langle x^2 g_{i1} \rangle}{dt}=  2 k_x \langle B \rangle \langle x g_{i1} \rangle  + k p_i\left \langle \sum_{j=1}^n  \left(\frac{x^2}{4}+\frac{x^2}{4}g_{i1}-x^2g_{i1}\right) g_{jj} \right \rangle  - k\langle x^2g_{i1} \rangle, \label{x2gi1}\\
&\frac{d\langle x^2 g_{ij} \rangle}{dt}=2 k_x \langle B \rangle \langle x g_{ij} \rangle 
- k\langle x^2 g_{ij} \rangle + k\langle x^2 g_{(i-1)j} \rangle, \ j=\left\lbrace 2,\ldots,i \right\rbrace, \label{x2gij}
\end{align} 
\end{subequations}
where dynamics of $\langle x^2 g_{i1} \rangle$ can be shown to follow
\begin{equation}
\frac{d\langle x^2 g_{i1} \rangle}{dt}=  2 k_x \langle B \rangle \langle x g_{i1} \rangle  + \frac{k}{4} p_i\left \langle \sum_{j=1}^n  {x^2}g_{jj} \right \rangle - \frac{3k}{4} p_i\left \langle \sum_{j=1}^n  x^2g_{i1} g_{jj} \right \rangle  - k\langle x^2g_{i1} \rangle.\label{x2gi11}
\end{equation}
Based on equation (10) in the main text $\left \langle \sum_{j=1}^n  x^2 g_{i1} g_{jj} \right\rangle =0$, thus equation \eqref{x2gi11} simplifies to
\begin{equation}
\frac{d\langle x^2 g_{i1} \rangle}{dt}=  2 k_x \langle B \rangle \langle x g_{i1} \rangle  + \frac{k}{4} p_i\left \langle \sum_{j=1}^n  {x^2}g_{jj} \right \rangle - k\langle x^2g_{i1} \rangle.\label{x2gi111}
\end{equation}
Equations \eqref{x2g11}, \eqref{x2gij}, and \eqref{x2gi111} can be compactly written as equation (16) in the main text.

\section{Moment dynamics of hybrid model introduced in Figure 2C}
Stochastic hybrid system introduced in Figure 2C coupled with phase-type distribution contains the following stochastic events 
\begin{figure}[h]
\centering
\includegraphics[width=0.8\textwidth]{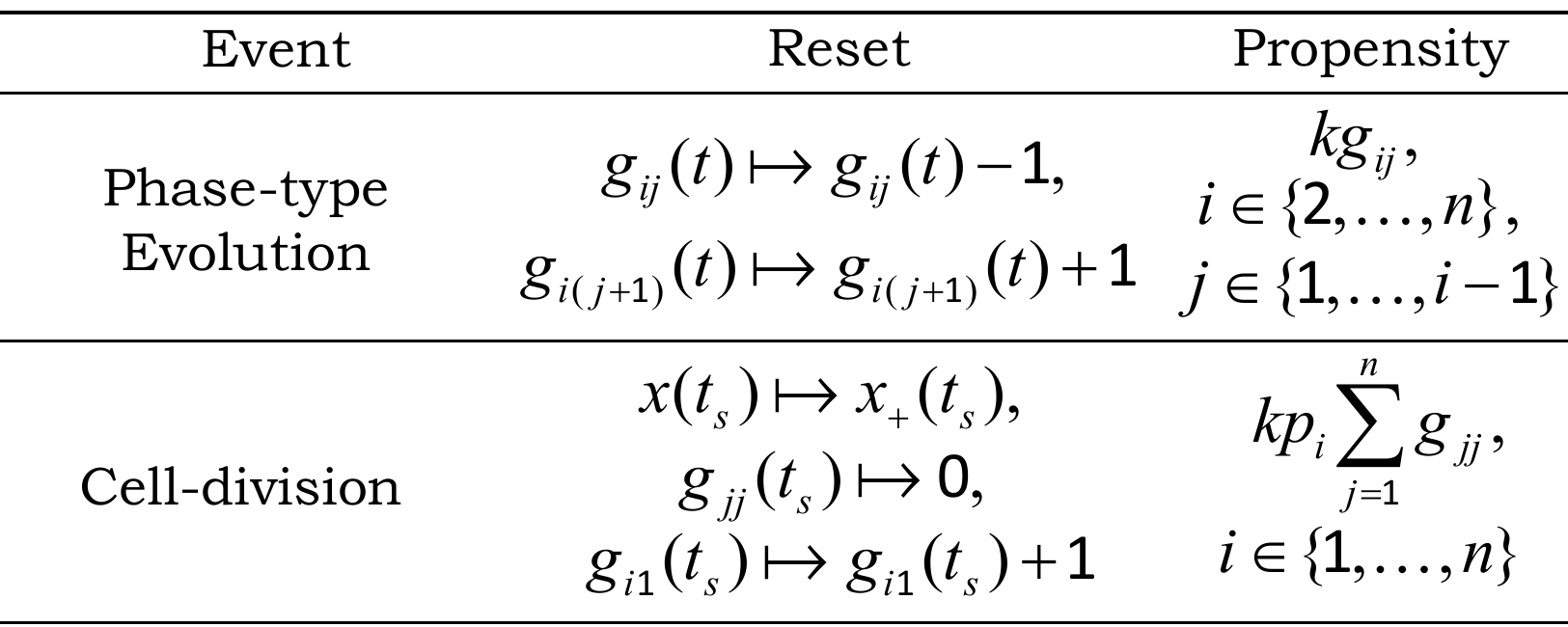} 
\end{figure}\\
and deterministic protein production dynamics
\begin{equation}
\dot{x}= k_x \langle B \rangle.
\end{equation}
Note that in this model $x(t)$ is a continuous random variable, thus we also use a continuous distribution to describe $x_+(t_s)$, however statistical properties of $x_+(t_s)$ is still given by (5). For this model we still can use equation \eqref{dynnf} to derive moment dynamics; equations describing time evolution of mean and $\langle xg_{ij}\rangle$ are the same as previous models, thus mean of protein for this model is equal to its value in Appendix A. The second order moment dynamics of protein can be written by choosing $\varphi$ to be $x^2$ in equation \eqref{dynnf}
\begin{equation}
\frac{d\langle x^2 \rangle}{dt}=
 2 k_x \langle B \rangle \langle x \rangle   +k \left \langle \sum_{j=1}^n  \left( \frac{x^2}{4}+ \frac{\alpha x}{4}-x^2 \right) g_{jj} \right \rangle ,  
\label{dynnf000}
\end{equation}
where conditional expected value of $x^2_+$ is substituted based on equation (5). Dynamics of $\langle x^2 \rangle$ can be simplified as
\begin{equation}
\frac{d\langle x^2 \rangle}{dt}=
 2 k_x \langle B \rangle \langle x \rangle  + \frac{ \alpha k}{4} \left \langle \sum_{j=1}^n  x g_{jj} \right \rangle - \frac{3k}{4}\left \langle \sum_{j=1}^n  x^2 g_{jj} \right \rangle.
\label{partitioning equations000}
\end{equation}
The same as before we add dynamics of the form $ \langle x^2g_{ij} \rangle $ to have a closed set of dynamics. First we add dynamics of $\langle x^2g_{11} \rangle$
\begin{equation}
\frac{d\langle x^2 g_{11} \rangle}{dt}=  2 k_x \langle B \rangle \langle x g_{11} \rangle  + \frac{\alpha k}{4} p_1 \left \langle  x g_{11}^2  \right \rangle  + \frac{k}{4} p_1 \left \langle  x^2 g_{11}^2  \right \rangle - k p_1 \left \langle  x^2 g_{11}^2  \right \rangle - k \sum_{i=2}^{n} p_i\langle x^2g_{11} \rangle, \label{x2g110}\\
\end{equation}
Based on equation (9) of the main text dynamics of $\langle x^2g_{11} \rangle$ simplifies to
\begin{equation}
\begin{aligned}
\frac{d\langle x^2 g_{11} \rangle}{dt}=  2 k_x \langle B \rangle \langle x g_{11} \rangle +\frac{\alpha k}{4} p_1 \left \langle  x g_{11}  \right \rangle  + \frac{k}{4} p_1 \left \langle  x^2 g_{11} \right \rangle  - k \langle x^2g_{11} \rangle. \label{x2g110}
\end{aligned}
\end{equation}
Now we express dynamics of moments $\langle x^2 g_{ij} \rangle$ for $g_{ij}\neq g_{11}$
\begin{subequations}
\begin{align}
&\frac{d\langle x^2 g_{i1} \rangle}{dt}=  2 k_x \langle B \rangle \langle x g_{i1} \rangle  + k p_i\left \langle \sum_{j=1}^n  \left(\frac{x^2}{4}+\frac{x^2}{4}g_{i1}+ \frac{\alpha x}{4}+ \frac{\alpha x}{4}g_{i1} -x^2g_{i1}\right) g_{jj} \right \rangle  - k\langle x^2g_{i1} \rangle, \label{x2gi10}\\
&\frac{d\langle x^2 g_{ij} \rangle}{dt}=2 k_x \langle B \rangle \langle x g_{ij} \rangle 
- k\langle x^2 g_{ij} \rangle + k\langle x^2 g_{(i-1)j} \rangle, \ j=\left\lbrace 2,\ldots,i \right\rbrace, \label{x2gij0}
\end{align} 
\end{subequations}
where dynamics of $\langle x^2 g_{i1} \rangle$ can be shown as
\begin{equation}
\begin{aligned}
\frac{d\langle x^2 g_{i1} \rangle}{dt}=&  2 k_x \langle B \rangle \langle x g_{i1} \rangle 
 + \frac{\alpha k}{4} p_i\left \langle \sum_{j=1}^n  {x}g_{jj} \right \rangle
 + \frac{k}{4} p_i\left \langle \sum_{j=1}^n  {x^2}g_{jj} \right \rangle  \\
&+ \frac{\alpha k}{4} p_i\left \langle \sum_{j=1}^n  xg_{i1} g_{jj} \right \rangle - \frac{3k}{4} p_i\left \langle \sum_{j=1}^n  x^2g_{i1} g_{jj} \right \rangle  - k\langle x^2g_{i1} \rangle.\label{x2gi110}
 \end{aligned}
\end{equation}
Based on equation (10) in the main text $\left \langle \sum_{j=1}^n  x^2 g_{i1} g_{jj} \right\rangle =0$, and  $\left \langle \sum_{j=1}^n  x g_{i1} g_{jj} \right\rangle =0$, hence equation \eqref{x2gi110} simplifies to
\begin{equation}
\frac{d\langle x^2 g_{i1} \rangle}{dt}=  2 k_x \langle B \rangle \langle x g_{i1} \rangle
 + \frac{\alpha k}{4} p_i\left \langle \sum_{j=1}^n  {x}g_{jj} \right \rangle
  + \frac{k}{4} p_i\left \langle \sum_{j=1}^n  {x^2}g_{jj} \right \rangle - k\langle x^2g_{i1} \rangle.\label{x2gi1110}
\end{equation}
Equations \eqref{partitioning equations000}, \eqref{x2g110}, \eqref{x2gij0}, and \eqref{x2gi1110} can be compactly written as equation (23) in the main text.

\section{Second and third-order moment dynamics of the full model}
Based on model introduced in Appendix A, second order moment dynamics of protein is expressed by choosing $\varphi$ to be $x^2$ in equation \eqref{general dynamic}, 
\begin{equation}
\frac{d\langle x^2 \rangle}{dt}=
 k_x \langle B^2 \rangle + 2 k_x \langle B \rangle \langle x \rangle   +k \left \langle \sum_{j=1}^n  \left( \frac{x^2}{4}+ \frac{\alpha x}{4}-x^2 \right) g_{jj} \right \rangle ,  
\label{dynnf0000}
\end{equation}
where conditional expected value of $x^2_+$ is substituted based on equation (5). Dynamics of $\langle x^2 \rangle$ can be simplified as
\begin{equation}
\frac{d\langle x^2 \rangle}{dt}=
  k_x \langle B^2 \rangle + 2 k_x \langle B \rangle \langle x \rangle  + \frac{ \alpha k}{4} \left \langle \sum_{j=1}^n  x g_{jj} \right \rangle - \frac{3k}{4}\left \langle \sum_{j=1}^n  x^2 g_{jj} \right \rangle.
\label{partitioning equations0000}
\end{equation}
The same as before we add dynamics of the form $ \langle x^2g_{ij} \rangle $ to have a closed set of moments. First we write dynamics of $\langle x^2 g_{11} \rangle $
\begin{equation}
\frac{d\langle x^2 g_{11} \rangle}{dt}= k_x \langle B^2 \rangle p_1+ 2 k_x \langle B \rangle \langle x g_{11} \rangle  + \frac{\alpha k}{4} p_1 \left \langle  x g_{11}^2  \right \rangle  + \frac{k}{4} p_1 \left \langle  x^2 g_{11}^2  \right \rangle - k p_1 \left \langle  x^2 g_{11}^2  \right \rangle - k \sum_{i=2}^{n} p_i\langle x^2g_{11} \rangle, \label{x2g1100}\\
\end{equation}
Based on equation (9) of the main text dynamics of $\langle x^2g_{11} \rangle$ simplifies to
\begin{equation}
\begin{aligned}
\frac{d\langle x^2 g_{11} \rangle}{dt}= k_x \langle B^2 \rangle p_1+ 2 k_x \langle B \rangle \langle x g_{11} \rangle +\frac{\alpha k}{4} p_1 \left \langle  x g_{11}  \right \rangle  + \frac{k}{4} p_1 \left \langle  x^2 g_{11} \right \rangle  - k \langle x^2g_{11} \rangle. \label{x2g1100}
\end{aligned}
\end{equation}
Next, dynamics of moments $\langle x^2 g_{ij} \rangle$ when $g_{ij}\neq g_{11}$ can be written as
\begin{subequations}
\begin{align}
\frac{d\langle x^2 g_{i1} \rangle}{dt}= & \frac{k_x \langle B^2 \rangle p_i}{i} +  2 k_x \langle B \rangle \langle x g_{i1} \rangle  \nonumber \\
&+ k p_i\left \langle \sum_{j=1}^n  \left(\frac{x^2}{4}+\frac{x^2}{4}g_{i1}+ \frac{\alpha x}{4}+ \frac{\alpha x}{4}g_{i1} -x^2g_{i1}\right) g_{jj} \right \rangle  - k\langle x^2g_{i1} \rangle, \label{x2gi10}\\
\frac{d\langle x^2 g_{ij} \rangle}{dt} =& \frac{k_x \langle B^2 \rangle p_i}{i} + 2 k_x \langle B \rangle \langle x g_{ij} \rangle 
- k\langle x^2 g_{ij} \rangle + k\langle x^2 g_{(i-1)j} \rangle, \ j=\left\lbrace 2,\ldots,i \right\rbrace, \label{x2gij00}
\end{align} 
\end{subequations}
where dynamics of $\langle x^2 g_{i1} \rangle$ can be shown as
\begin{equation}
\begin{aligned}
\frac{d\langle x^2 g_{i1} \rangle}{dt}=&  \frac{k_x \langle B^2 \rangle p_i}{i} + 2 k_x \langle B \rangle \langle x g_{i1} \rangle 
 + \frac{\alpha k}{4} p_i\left \langle \sum_{j=1}^n  {x}g_{jj} \right \rangle
 + \frac{k}{4} p_i\left \langle \sum_{j=1}^n  {x^2}g_{jj} \right \rangle  \\
&+ \frac{\alpha k}{4} p_i\left \langle \sum_{j=1}^n  xg_{i1} g_{jj} \right \rangle - \frac{3k}{4} p_i\left \langle \sum_{j=1}^n  x^2g_{i1} g_{jj} \right \rangle  - k\langle x^2g_{i1} \rangle.\label{x2gi1100}
 \end{aligned}
\end{equation}
Based on equation (10) in the main text $\left \langle \sum_{j=1}^n  x^2 g_{i1} g_{jj} \right\rangle =0$ and  $\left \langle \sum_{j=1}^n  x g_{i1} g_{jj} \right\rangle =0$, hence equation \eqref{x2gi1100} simplifies to
\begin{equation}
\frac{d\langle x^2 g_{i1} \rangle}{dt}=  \frac{k_x \langle B^2 \rangle p_i}{i} +  2 k_x \langle B \rangle \langle x g_{i1} \rangle
 + \frac{\alpha k}{4} p_i\left \langle \sum_{j=1}^n  {x}g_{jj} \right \rangle
  + \frac{k}{4} p_i\left \langle \sum_{j=1}^n  {x^2}g_{jj} \right \rangle - k\langle x^2g_{i1} \rangle.\label{x2gi11100}
\end{equation}
Equations \eqref{partitioning equations0000}, \eqref{x2g1100}, \eqref{x2gij00}, and \eqref{x2gi11100} can be compactly written as equation (26) in the main text.

\section{Contribution of different sources of stochasticity in protein by taking into account gene-duplication}

\begin{figure}[h]
\centering
\includegraphics[width=0.75\textwidth]{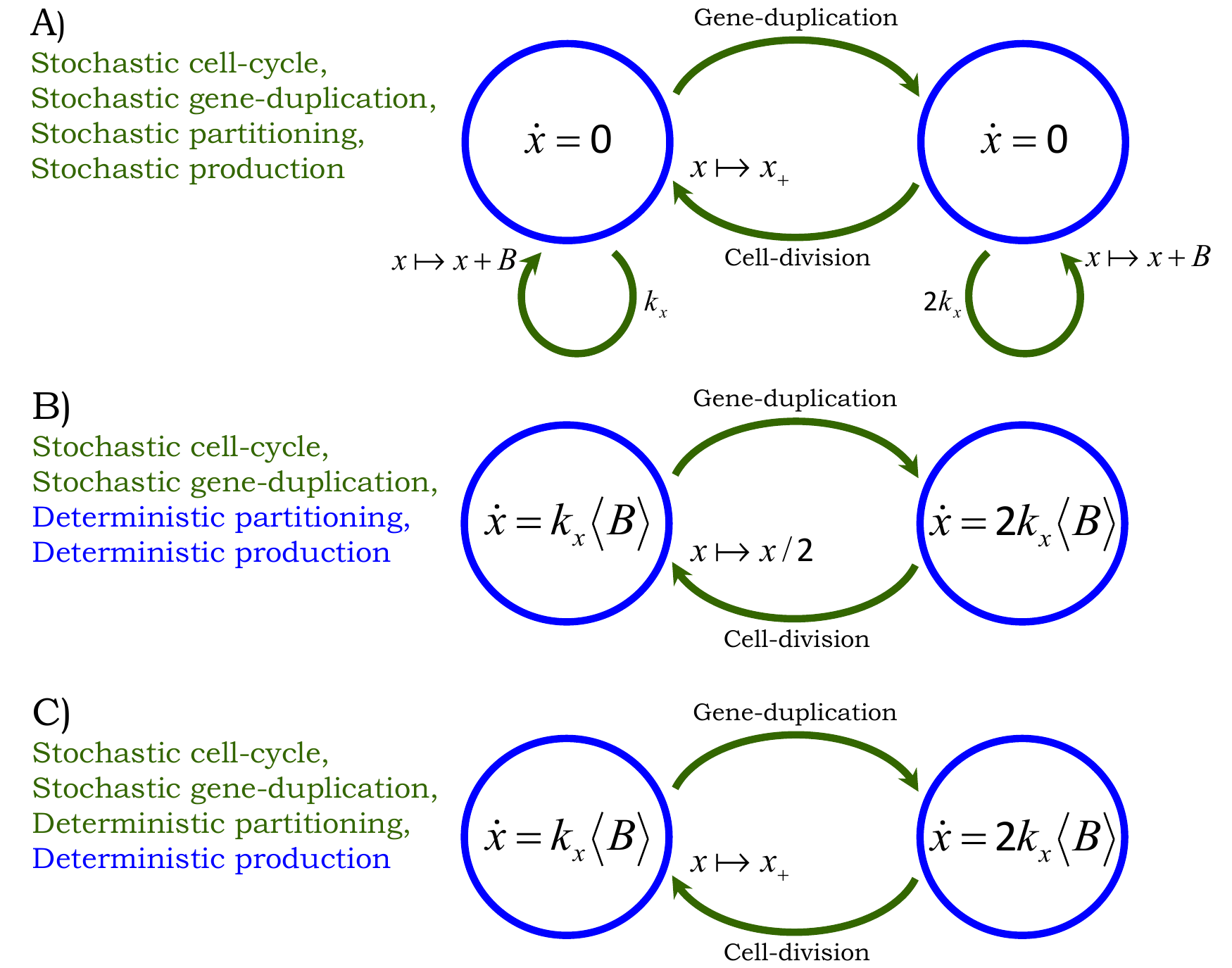} 
\caption*{Figure S1: {\bf Stochastic hybrid models for quantifying different sources of noise.}  Gene-duplication and cell-division times are random events. \textbf{A)} Protein production happens in random bursts with burst frequency $k_x$. After gene-duplication event burst frequency doubles ($2k_x$). In the time of division proteins will be distributed between mother and daughter cells randomly, and the protein burst frequency will be $k_x$ again. \textbf{B)} Protein production is considered in a deterministic fashion, and after gene-duplication dynamics of protein production is multiplied by a factor of two, i.e., $\dot{x}=2k_x \langle B \rangle$. In the division event proteins are distributed between mother and daughter cells equally. Thus the only stochastic events are duplication and division events. \textbf{C)} Protein is produced in a deterministic fashion, but in time of division protein levels in daughter and mother cells are random. Thus duplication, division, and partitioning are random events.
}
\label{fig:target distribution}
\end{figure}

\begin{figure}[!h]
\center
\includegraphics[width=0.66\textwidth]{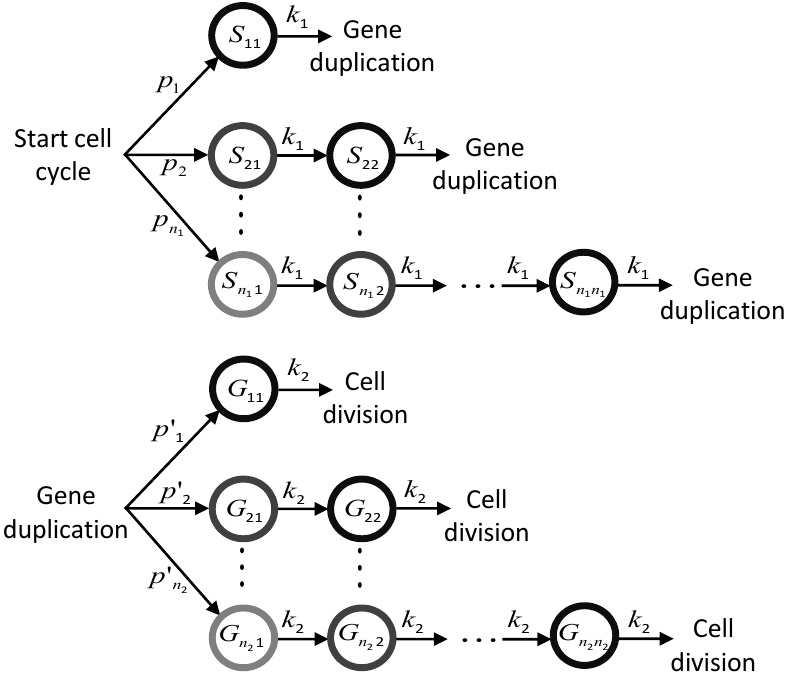}
\caption*{Figure S2: {\bf Cell-cycle time consists of two time intervals: at the end of the first interval gene duplicates, and at the end of the second one cell divides.} Two independent phase-type distributions are used to model cell-cycle time in the presence of genome duplication. The states of the first distribution are denoted by $S_{ij} , \ i=\{1,\ldots,n_1\}, \ j=\{1,\ldots,i\}$; transition between these states happens at a constant rate $k_1$. The states of the second distribution are shown by $G_{ij} , \ i=\{1,\ldots,n_2\}, \ j=\{1,\ldots,i\}$, and transition between these states occurs at a rate $k_2$.
}
\label{fig:target distribution}
\end{figure}

We study the contribution of different sources of stochasticity by using models introduced in Figure S1. The cell-cycle time consists of two time intervals: the time interval before gene-duplication and the time after gene-duplication. These time intervals are modeled by using two independent phase-type distributions as shown in Figure S2. 
Based on phase-type characteristics mean of the states of the first phase-type $\langle s_{ij} \rangle $ and the second phase-type $\langle g_{ij} \rangle $ are
\begin{equation}
\begin{aligned}
&\langle s_{ij} \rangle = \frac{p_i}{i}\beta,\ \ i\in\{1,\ldots,n_1\}, \ \ j\in \{1,\ldots,i \}, \\
& \langle g_{ij} \rangle = \frac{p'_i}{i}(1-\beta), \ \ i\in\{1,\ldots,n_2\}, \ \ j\in \{1,\ldots,i \},
\end{aligned}
\label{phase mean}
\end{equation}
where $\beta$ is defined as
\begin{equation}
\beta \coloneqq \frac{\text{Mean time interval before gene-duplication}}{\text{Mean cell-cycle time}}=\frac{\langle T_1 \rangle }{\langle T \rangle }. \label{a eq.}
\end{equation}
We start our analysis by deriving mean level of protein in the next section.

\subsection{Mean of protein count level in the presence of gene-duplication}
After gene-duplication the amount of genes expressing a specific protein doubles. Thus the rate of protein production increases by a factor of two as shown in Figure S1A. This model coupled with phase-type distributions contains the following stochastic events 
\begin{figure}[h]
\centering
\includegraphics[width=0.8\textwidth]{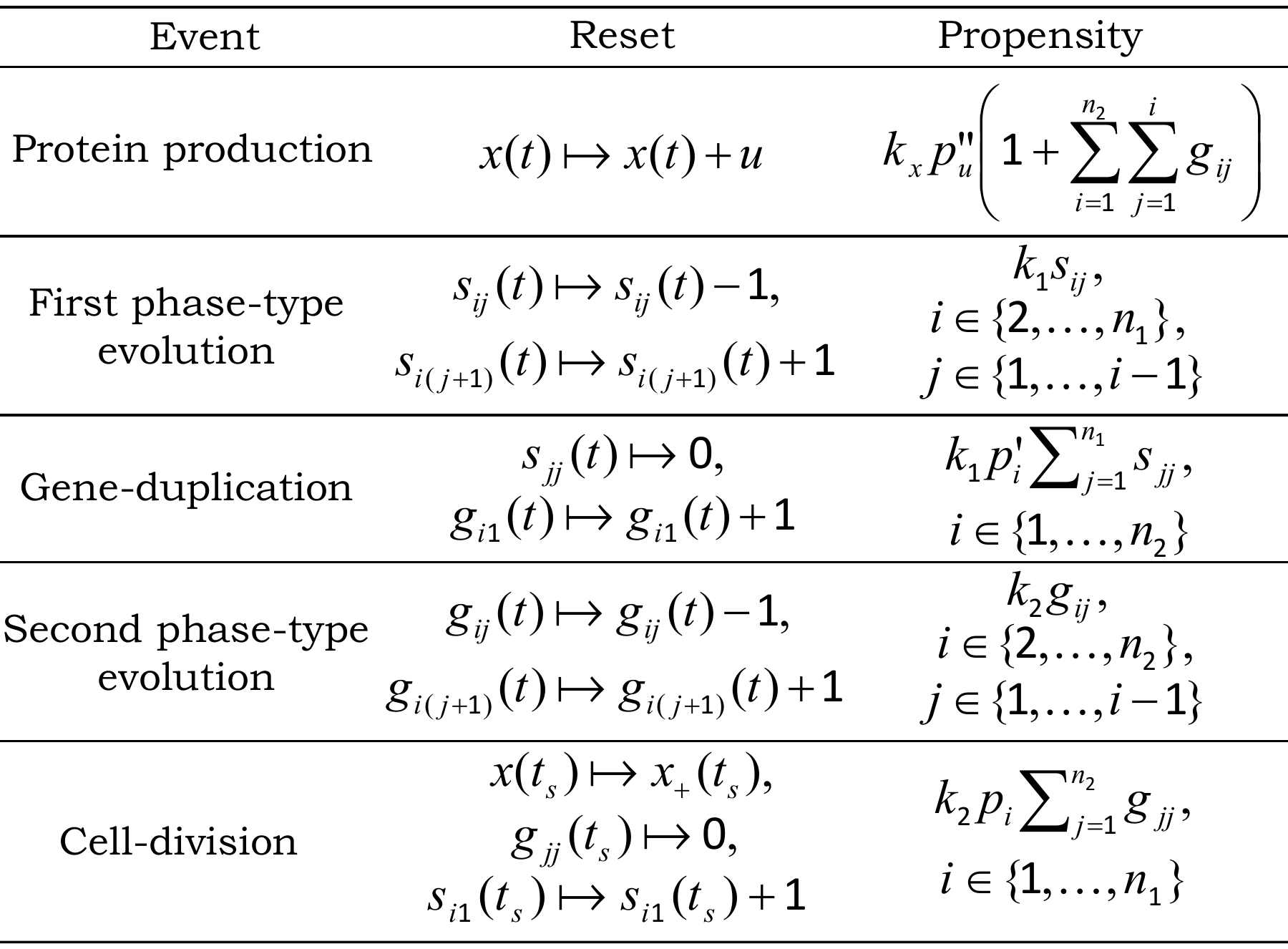} 
\end{figure}\\
Note that in the protein production event, before gene-duplication all the states $g_{ij}$ are zero thus propensity function will be $k_x p''_u$. After gene-duplication and before division, one of the states $g_{ij}$ is one hence propensity function will be $2k_x p''_u$. In time of gene-duplication, states of the first phase-type will reset to zero and state $g_{i1}$ of the second distribution will be selected with probability $p'_i$; hence propensity function of gene-duplication event is $ k_1 p'_i \sum_{j=1}^{n_1}s_{jj}$. At the end of cell-cycle, states of the second phase-type will reset to zero and a new cell-cycle which is sum of $i$ exponentials will be selected with probability $p_i$; thus propensity function of cell-division event is $ k_2 p_i \sum_{j=1}^{n_1}g_{jj}$.

Theorem 1 of \cite{hsi04} gives the time derivative of the expected value of any function $\varphi (x,s_{ij},g_{ij})$ as 
\begin{equation}
\frac{d\langle \varphi (x,s_{ij},g_{ij}) \rangle}{dt}= \left \langle \sum_{Events}  \Delta \varphi (x,s_{ij},g_{ij}) \times f(x,s_{ij},g_{ij}) \right \rangle, 
\label{general dynamic1}  
\end{equation}
where $\Delta \varphi(x,s_{ij},g_{ij})$ is a change in $\varphi$ when an event occurs. The first-order moment dynamic of this model can be expressed by selecting $\varphi$ to be $x$ in equation \eqref{general dynamic1}
\begin{equation}
\begin{aligned}
\frac{d\langle x \rangle}{dt}=& k_x \langle B \rangle \left( 1+\left  \langle  \sum_{i=1}^{n_2}  \sum_{j=1}^i  g_{ij} \right \rangle\right)   - k_2\left  \langle  \sum_{j=1}^{n_2}  (\frac{x}{2} -x) g_{jj} \right \rangle,\\
\end{aligned} 
\end{equation}
where conditional expected value of $x_+$ is replaced from equation (5); by using equation \eqref{phase mean} mean dynamics can be simplified as
\begin{equation}
\begin{aligned}
\frac{d\langle x \rangle}{dt}=& k_x \langle B \rangle \left( 2-\beta \right)   -\frac{k_2}{2}\left  \langle  \sum_{j=1}^{n_2}  x g_{jj} \right \rangle,\\
\end{aligned} \label{mean1}
\end{equation}

Mean dynamics is not closed thus we add dynamics of $\langle x s_{ij} \rangle, \ i=\{1,\ldots,n_1\}, \ j=\{1,\ldots,i\}$ and $\langle x g_{ij} \rangle, \ i=\{1,\ldots,n_1\}, \ j=\{1,\ldots,i\}$ to have a closed set of moment equations. These moment dynamics are simplified by using equations (5), (9), (10) and \eqref{phase mean} as
\begin{subequations}
\begin{align}
& \frac{d\langle x s_{i1} \rangle}{dt}= \frac{k_x \langle B \rangle p'_i \beta}{i}  + \frac{k_2}{2}p'_i\left \langle \sum_{j=1}^{n_2} x g_{jj} \right \rangle - k_1\langle xs_{i1} \rangle,\label{genes1}\\
&\frac{d\langle x s_{ij} \rangle}{dt}= \frac{k_x \langle B \rangle p'_i \beta}{i}
 - k_1\langle x s_{ij} \rangle + k_1\langle x s_{i (j-1)} \rangle, \ 
\ j= \lbrace 2,\ldots,i \rbrace,\label{genes2}\\
& \frac{d\langle x g_{i1} \rangle}{dt}= \frac{ 2 k_x \langle B \rangle p_i (1-\beta)}{i} + k_1 p_i\left \langle \sum_{j=1}^{n_1} x s_{jj} \right \rangle - k_2\langle xg_{i1} \rangle, \label{gene1}\\
&\frac{d\langle x g_{ij} \rangle}{dt}= \frac{ 2 k_x \langle B \rangle p_i (1-\beta )}{i} - k_2\langle x g_{ij} \rangle + k_2\langle x g_{i (j-1)} \rangle,\ j= \lbrace 2,\ldots,i \rbrace.\label{gene2}
\end{align} 
\end{subequations}

In order to find the mean of protein, first we need to find the moments $\overline{\langle x s_{ij} \rangle}, \ i=\left\lbrace 1,\ldots,n_1 \right\rbrace, \ j= \left\lbrace 1,\ldots,i \right\rbrace $ and $\overline{\langle x g_{ij}\rangle}, \ i=\left\lbrace 1,\ldots,n_2 \right\rbrace, \ j= \left\lbrace 1,\ldots,i \right\rbrace $. For calculating these moments we should calculate the term $ \overline{ \left \langle  \sum_{j=1}^{n_2}  x g_{jj} \right \rangle}$; this term can be obtained by analyzing equation \eqref{mean1} in steady-state
\begin{equation}
\begin{aligned}
k_x\langle B \rangle (2-\beta)=   \frac{k_2}{2}\overline{ \left \langle  \sum_{j=1}^{n_2} x g_{jj} \right \rangle }\Rightarrow \overline{ \left \langle  \sum_{j=1}^{n_2}  x g_{jj} \right \rangle}= \frac{2 k_x\langle B \rangle (2-\beta)}{k_2}.\label{last gene1}
\end{aligned} 
\end{equation}
By having this term, we calculate $\overline{\langle x s_{ij} \rangle}$ by recursion process: we start by calculating $\overline{ \langle x s_{i1} \rangle}$ by substituting equation \eqref{last gene1} in equation \eqref{genes1}. In the next step we use the definition we derived for $\overline{\langle x s_{i1} \rangle}$ to calculate $\overline{\langle x s_{i2} \rangle}$ from equation \eqref{genes2}. We continue this process until we derive all the moments
\begin{equation}
\begin{aligned}
 \overline{  \langle x s_{ij} \rangle}= \frac{k_x\langle B \rangle}{k_1}p'_i\left( \beta \frac{j}{i} + (2-\beta)\right), \ i= \lbrace 1,\ldots,n_1 \rbrace, \ j= \lbrace 1,\ldots,i \rbrace. 
  \label{all gene s extrinsic}
\end{aligned} 
\end{equation}
Now we need to calculate the moments $\overline{ \langle x g_{ij} \rangle}, \ i= \left\lbrace 1,\ldots,n_2 \right\rbrace, \ j= \left\lbrace 1,\ldots,i \right\rbrace $, thus we need the expression of the term $\overline{\left \langle  \sum_{j=1}^{n_1}  x s_{jj} \right \rangle}$; from equation \eqref{all gene s extrinsic} we have the following 
\begin{equation}
\begin{aligned}
\overline{\left \langle  \sum_{j=1}^{n_1}  x s_{jj} \right \rangle}= \frac{2 k_x\langle B \rangle}{k_1}.\label{last gene s1}
\end{aligned} 
\end{equation}
Substituting this term in equations \eqref{gene1} and \eqref{gene2} result in 
\begin{equation}
\begin{aligned}
\overline{ \langle x g_{ij} \rangle}= \frac{2k_x\langle B \rangle}{k_2}p_i\left( (1-\beta) \frac{j}{i} + 1\right), \  i=\lbrace 1,\ldots,n_2 \rbrace, \ j= \lbrace 1,\ldots,i \rbrace. 
  \label{all gene}
\end{aligned} 
\end{equation}
Note that 
\begin{equation}
\begin{aligned}
\sum_{i=1}^{n_1}\sum_{j=1}^{i}  s_{ij}+ \sum_{i=1}^{n_2}\sum_{j=1}^{i} g_{ij} = 1 &\Rightarrow   \langle x \rangle = \left \langle x \left(\sum_{i=1}^{n_1}\sum_{j=1}^{i}  s_{ij}+ \sum_{i=1}^{n_2}\sum_{j=1}^{i} g_{ij} \right)\right \rangle 
\\
 &\Rightarrow  \overline{ \langle x \rangle} =\sum_{i=1}^{n_1}\sum_{j=1}^{i}\overline{ \langle  x s_{ij}\rangle }+ \sum_{i=1}^{n_2}\sum_{j=1}^{i} \overline{ \langle  x g_{ij} \rangle}.
\end{aligned}
\end{equation}
Thus by adding all the term calculated here and using equation (7) mean of protein can be calculated as 
\begin{equation}
\begin{aligned}
\overline{ \langle x \rangle}=\frac{k_x \langle B \rangle \langle {T_1} \rangle \left(4-\beta +  \beta CV^2_{T_1} \right)}{2}+ k_x \langle B \rangle \langle {T_2} \rangle \left(3-\beta +  (1-\beta) CV^2_{T_2} \right)\label{mean00}.
\end{aligned}
\end{equation}

\newpage
\subsection{Noise in protein count level contributed from cell-cycle time}
In order to calculate the noise contributed from cell-cycle time variation, the model introduced in Figure S1B coupled with phase-type distributions is used. This model contains following stochastic events 
\begin{figure}[h]
\centering
\includegraphics[width=0.8\textwidth]{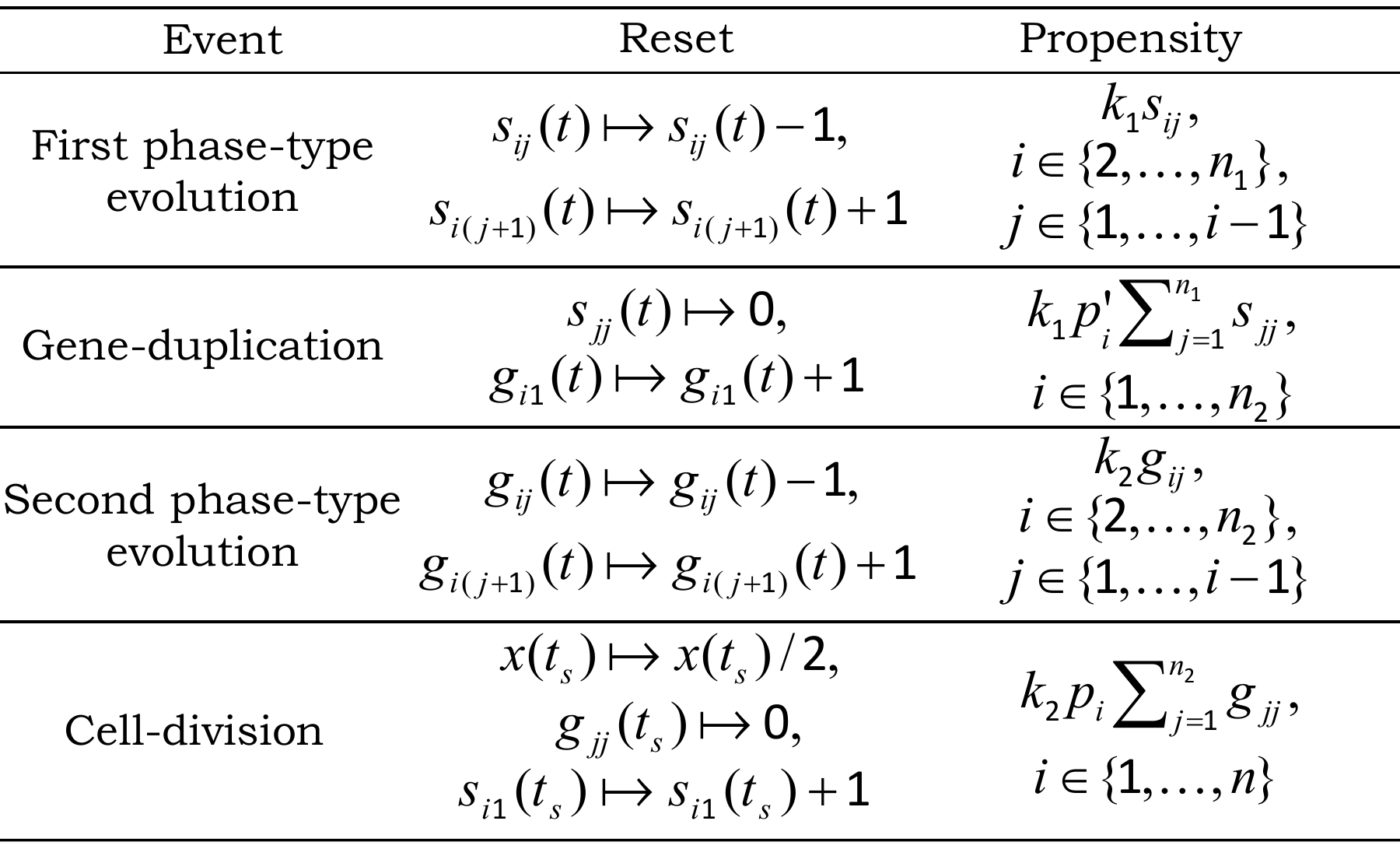} 
\end{figure}\\
and deterministic protein production
\begin{equation}
\dot{x}= k_x \langle B \rangle\left(1+ \sum_{i=1}^{n_2}\sum_{j=1}^{i}g_{ij} \right).
\end{equation}

Theorem 1 of \cite{hsi04} gives the time derivative of the expected value of any function $\varphi (x,s_{ij},g_{ij})$ as 
\begin{equation}
\begin{aligned}
\frac{d\langle \varphi (x,s_{ij},g_{ij}) \rangle}{dt}= &\left \langle \sum_{Events}  \Delta \varphi (x,s_{ij},g_{ij}) \times f(x,s_{ij},g_{ij}) \right \rangle \\
&+ \left \langle \frac{\partial \varphi(x,g_{ij})}{\partial x} k_x \langle B \rangle\left(1+ \sum_{i=1}^{n_2}\sum_{j=1}^{i}g_{ij}\right) \right \rangle , \label{general dynamic2}  
\end{aligned}
\end{equation}
where the first term in the right hand side is contributed from stochastic events, and the second term is contributed from deterministic protein production. In this model, dynamics of $\langle x\rangle$, $\langle xs_{ij}\rangle$ and $\langle xg_{ij}\rangle$ are the same as equations \eqref{mean1} and (E.6), thus mean of protein, $\overline{ \langle xs_{ij} \rangle}$, and $\overline{ \langle xg_{ij} \rangle}$ will be equal to their value in previous section. Further, the second-order moment dynamics of protein can be added by selecting $\varphi$ to be $x^2$ in equation \eqref{general dynamic2}
\begin{equation}
\begin{aligned}
&\frac{d\langle x^2 \rangle}{dt}=
2 k_x \langle B \rangle \left( \langle x \rangle+\left  \langle  \sum_{i=1}^{n_2}  \sum_{j=1}^i x g_{ij} \right \rangle\right)
- \frac{3 k_2}{4} \left \langle\sum_{j=1}^{n_2}  x^2 g_{jj}\right  \rangle.   \label{x21}
\end{aligned} 
\end{equation}
This equation is not closed thus we add dynamics of ${\langle x^2 s_{ij}} \rangle, \ i=\left\lbrace 1,\ldots,n_1 \right\rbrace, \ j= \left\lbrace 1,\ldots,i \right\rbrace $ and ${\langle x^2 g_{ij}} \rangle, \ i=\left\lbrace 1,\ldots,n_2 \right\rbrace, \ j= \left\lbrace 1,\ldots,i \right\rbrace $ to have a closed set of equations
\begin{subequations}
\begin{align}
&\frac{d\langle x^2 s_{i1} \rangle}{dt}=  2 k_x \langle B \rangle \langle x s_{i1} \rangle  + \frac{k_2}{4} p_i\left \langle \sum_{j=1}^{n_2}  x^2 g_{jj}\right  \rangle  - k_1\langle x^2s_{i1} \rangle,  \label{higher_moment_partition s1}\\
&\frac{d\langle x^2 s_{ij} \rangle}{dt}=2 k_x \langle B \rangle \langle x s_{ij} \rangle 
- k_1\langle x^2 s_{ij} \rangle + k_1\langle x^2 s_{(i-1)j} \rangle, \ j= \left\lbrace 2,\ldots,i \right\rbrace,
\label{higher_moment_partition s2}\\
&\frac{d\langle x^2 g_{i1} \rangle}{dt}=  4 k_x \langle B \rangle \langle x g_{i1} \rangle  + k_1 p_i \left \langle \sum_{j=1}^{n_1}  x^2 s_{jj} \right \rangle  - k_2\langle x^2g_{i1} \rangle,\label{higher_moment_partition1}\\
&\frac{d\langle x^2 g_{ij} \rangle}{dt}=4 k_x \langle B \rangle \langle x g_{ij} \rangle 
- k_2 \langle x^2 g_{ij} \rangle + k_2 \langle x^2 g_{(i-1)j} \rangle,  j= \left\lbrace 2,\ldots,i \right\rbrace.
\label{higher_moment_partition2}
\end{align} 
\end{subequations}
In order to calculate noise we need to express $\overline{\langle x^2s_{ij} \rangle}$, and $\overline{ \left \langle x^2g_{ij} \right \rangle}$, which requires calculating the term $\overline{ \langle  \sum_{j=1}^{n_2}  x^2 g_{jj} \rangle}$; this term can be derived by analyzing equation \eqref{x21} in steady-state
\begin{equation}
\begin{aligned}
& \frac{3 k_2}{4} \overline{\left \langle\sum_{j=1}^{n_2}  x^2 g_{jj} \right  \rangle } =2 k_x \langle B \rangle \left( \overline{\langle x \rangle}+\overline{ \left  \langle  \sum_{i=1}^{n_2}  \sum_{j=1}^i  x g_{ij} \right  \rangle}\right)  \Rightarrow \\
&
\overline{ \left \langle\sum_{j=1}^{n_2}  x^2 g_{jj} \right \rangle}  =\frac{4k_x^2 \langle B \rangle^2 \langle T_1 \rangle\left((4-\beta)+\beta CV^2_{T_1}\right)}{3k_2}+\frac{16 k_x^2 \langle B \rangle^2 \langle T_2\rangle \left((3-\beta)+(1-\beta) CV^2_{T_1}\right)}{3 k_2},\label{last gene}
\end{aligned} 
\end{equation}
where in deriving this term we used equation \eqref{mean00} and we summed all the terms in equation \eqref{all gene}. By having this term, we calculate $\overline{\langle x^2 s_{ij} \rangle}$ by recursion process. we derive $\overline{ \langle x^2 s_{i1} \rangle}$ by substituting equation \eqref{last gene} in equation \eqref{higher_moment_partition s1}. In the next step we use the definition of $\overline{\langle x^2 s_{i1} \rangle}$ to calculate $\overline{\langle x^2 s_{i2} \rangle}$ from equation \eqref{higher_moment_partition s2}. We continue this process until we derive all the moments
\begin{equation}
\begin{aligned}
\overline{ \langle x^2 s_{ij} \rangle}=&\frac{k_x^2 \langle B \rangle^2 \langle T_1 \rangle\left((4-\beta)+\beta CV^2_{T_1}\right)}{3k_1}p'_i+\frac{ 4k_x^2 \langle B \rangle^2 \langle T_2\rangle \left((3-\beta)+(1-\beta) CV^2_{T_2}\right)}{3 k_1}p'_i\\
&+\frac{2k_x^2 \langle B \rangle^2}{k_1}p'_i\left( \frac{ \beta j^2 +(2-\beta)j}{i} \right),  \ i= \{1,\ldots,n_1\}, \  j=\{1,\ldots,i\}.\label{all higher partition00}
\end{aligned} 
\end{equation}
Expressing $\overline{\langle x^2 g_{ij} \rangle}$ requires calculation of the term $\overline{ \langle  \sum_{j=1}^{n_1}  x^2 s_{jj} \rangle}$ which can be obtained from equation \eqref{all higher partition00} as
\begin{equation}
\begin{aligned}
\overline{ \left \langle \sum_{j=1}^{n_1}  x^2 s_{jj} \right \rangle}=\frac{4k_x^2 \langle B \rangle^2 \langle T_1 \rangle\left((4-\beta)+\beta CV^2_{T_1}\right)}{3k_1}+\frac{ 4k_x^2 \langle B \rangle^2 \langle T_2\rangle \left((3-\beta)+(1-\beta) CV^2_{T_2}\right)}{3 k_1}.\label{all higher partition}
\end{aligned} 
\end{equation}
Thus $\overline{ \langle x^2g_{ij} \rangle}$ can be obtained with a recursion process from equations \eqref{higher_moment_partition1} and \eqref{higher_moment_partition2}
\begin{equation}
\begin{aligned}
\overline{ \langle x^2g_{ij} \rangle}=&\frac{4 k_x^2 \langle B \rangle^2 \langle\langle T_1 \rangle\left((4-\beta)+\beta  CV^2_{T_1}\right)}{3k_2}p_i+\frac{ 4 k_x^2 \langle B \rangle^2 \langle T_2\rangle \left((3-\beta)+(1-\beta) CV^2_{T_2}\right)}{3 k_2}p_i\\
&
+\frac{8k_x^2 \langle B \rangle^2}{k_2}p_i\left( \frac{(1-\beta)j^2 +j}{i} \right),  \ i= \{1,\ldots,n_2\}, \ j= \{1,\ldots,i\}.\label{all higher partition}
\end{aligned} 
\end{equation}
Note that $ \sum_{i=1}^{n_1}\sum_{j=1}^{i}\overline{ \langle x^2 s_{ij} \rangle}+ \sum_{i=1}^{n_2}\sum_{j=1}^{i}\overline{\langle  x^2 g_{ij}\rangle} =\overline{ \langle x^2 \rangle}$ thus the second order moment of protein can be derived by adding all the terms in equations \eqref{all higher partition00} and \eqref{all higher partition}. $\overline{ \langle x^2 \rangle}$ can be simplified by using equations (7) and (18b) in the main article as
\begin{equation}
\begin{aligned}
&\overline{ \langle x^2 \rangle}  =
k_x^2 \langle B \rangle^2\frac{4 \langle T_1^3\rangle+16 \langle T_2^3 \rangle +2 \langle T\rangle^3 (3(2-\beta)^2+ \beta^2(5-3\beta) CV^2_{T_1} +8(1-\beta)^2  CV^2_{T_2})}{3 \langle T\rangle}.
  \end{aligned}
\end{equation}
Finally, using the definition of $CV^2$ results in noise of protein raised from cell-cycle time variations
\begin{footnotesize}
\begin{equation}
\begin{aligned}
&CV^2_E = \frac{(4 \langle T_1^3\rangle+16 \langle T_2^3\rangle)/\langle T \rangle^3-3 (2 - 4 \beta + \beta^2)^2  }{3\left((\beta^2 -4 \beta +6) + \beta^2CV^2_{T_1} + 2(1-\beta)^2 CV^2_{T_2}  \right)^2}\\
&\frac{ -3 \beta^2 ( \beta^2 (-2 + CV^2_{T_1})) CV^2_{T_1} -4 (\beta^2CV^2_{T_1}+(1-\beta)^2 CV^2_{T_2})(2 - 12 \beta + 3 \beta^2 + 3 (\beta^2CV^2_{T_1}+(1-\beta)^2 CV^2_{T_2})) }{3\left((\beta^2 -4 \beta +6) + \beta^2CV^2_{T_1} + 2(1-\beta)^2 CV^2_{T_2} \right)^2}
.\label{cv_division_time}
\end{aligned}
\end{equation}
\end{footnotesize}

\newpage

\subsection{Noise in protein count level contributed from random partitioning}

In order to take into account noise caused by random partitioning of proteins between two daughter cells, we use the model shown in Figure S1C coupled with phase-type distributions. This model contains the following stochastic events 
\begin{figure}[h]
\centering
\includegraphics[width=0.8\textwidth]{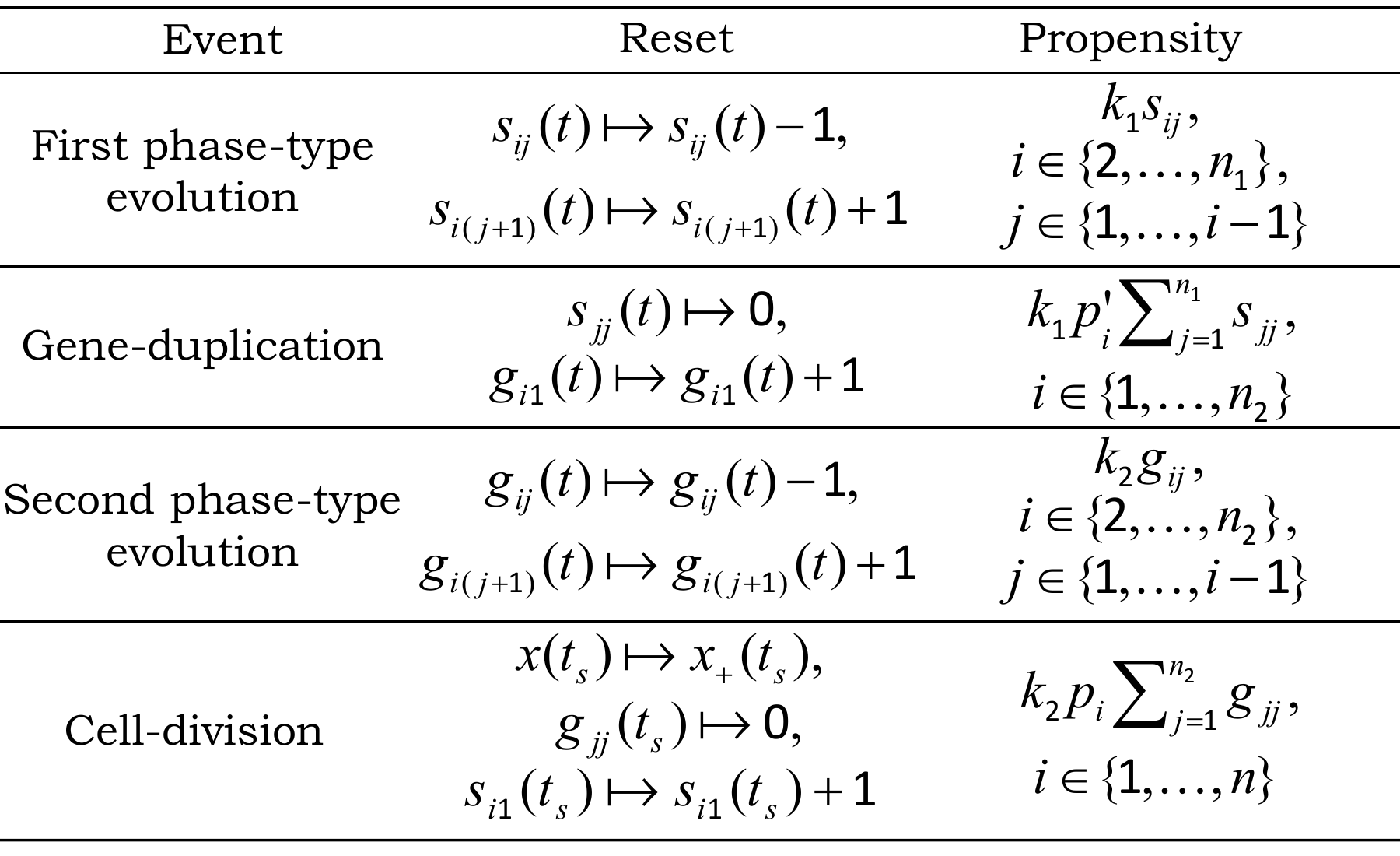} 
\end{figure}\\
and deterministic protein production
\begin{equation}
\dot{x}= k_x \langle B \rangle \sum_{i=1}^{n_2}\sum_{j=1}^{i}g_{ij}.
\end{equation}
Note that here $x$ is a continuous random variable, hence $x_+$ is also obtained from a continious distribution. Connection between statistical statistical moments of $x$ and $x_+$ is given by (5).

For this model, $\overline{\langle x \rangle}$, $\overline{\langle xs_{ij}\rangle}$, and $\overline{\langle xg_{ij}\rangle}$ are equal to their value in Section E.1 and Section E.2. 
However, dynamics of $\langle x^2 \rangle$ and $\langle x^2s_{i1}\rangle$ are different
\begin{subequations}
\begin{align}
&\frac{d\langle x^2 \rangle}{dt}=
2 k_x \langle B \rangle \left( \langle x \rangle+ \left \langle  \sum_{i=1}^{n_2}  \sum_{j=1}^i   x g_{ij}\right \rangle\right)
+ \frac{1}{4}\alpha k_2\left \langle\sum_{j=1}^{n_2} x  g_{jj} \right \rangle-\frac{3 k_2}{4}\left  \langle\sum_{j=1}^{n_2} x^2  g_{jj} \right\rangle, \label{x2 partirion} \\
&\frac{d\langle x^2 s_{i1} \rangle}{dt}=  2 k_x \langle B \rangle \langle x s_{i1} \rangle +\frac{ k_2}{4}   p'_i\left \langle \sum_{j=1}^{n_2}  x^2 g_{jj} \right \rangle + \frac{1}{4}\alpha k_2 p'_i\left \langle \sum_{j=1}^{n_2}  x g_{jj}\right  \rangle - k_1\langle x^2s_{i1} \rangle, \label{x2s1 partition}
 \end{align} 
\end{subequations}
note that dynamics of $\langle x^2 s_{ij} \rangle$, $\langle x^2 g_{i1} \rangle$, and  $\langle x^2 g_{ij} \rangle$ are identical to equations \eqref{higher_moment_partition s2}, \eqref{higher_moment_partition1}, and \eqref{higher_moment_partition2}. 
Similar to previous section, we start by deriving the term $\overline{ \langle\sum_{j=1}^n x^2  g_{jj} \rangle }$. Analyzing equation \eqref{x2 partirion} in steady-state gives this term as
\begin{equation}
\begin{aligned}
 \overline{ \left \langle\sum_{j=1}^{n_2}  x^2 g_{jj} \right \rangle}  = &  \frac{4k_x^2 \langle B \rangle^2 \langle T_1 \rangle\left((4-\beta )+\beta CV^2_{T_1}\right)}{3k_2}\\
&+\frac{16 k_x^2 \langle B \rangle^2 \langle T_2\rangle \left((3-\beta)+(1-\beta) CV^2_{T_1}\right)}{3 k_2}
 +\frac{2 \alpha k_x \langle B \rangle (2-\beta)}{3 k_2} .\label{last gene partitioning01}
\end{aligned}
\end{equation}
Substituting equation \eqref{last gene partitioning01} in equations \eqref{x2s1 partition} and \eqref{higher_moment_partition s2} results in 
\begin{equation}
\begin{aligned}
 \overline{  \langle x^2 s_{ij} \rangle}=
 &\frac{k_x^2 \langle B \rangle^2 \langle T_1 \rangle\left((4-\beta)+\beta CV^2_{T_1}\right)}{3k_1}p'_i+\frac{ 4k_x^2 \langle B \rangle^2 \langle T_2\rangle \left((3-\beta)+(1-\beta) CV^2_{T_2}\right)}{3 k_1}p'_i\\
&+\frac{2k_x^2 \langle B \rangle^2}{k_1}p'_i\left( \frac{ \beta j^2 +(2-\beta)j}{i} \right)+\frac{2 \alpha k_x(2-\beta)\langle B \rangle}{3k_1}p'_i, \\& i= \{1,\ldots,n_1\}   \ j= \{1,\ldots,i\}.\label{all higher partition011}
\end{aligned} 
\end{equation}
In the next step we derive moments $\overline{\langle x^2 g_{ij} \rangle}$; we start by calculating $ \overline{ \langle  \sum_{j=1}^{n_1}  x^2 s_{jj} \rangle}$ from \eqref{all higher partition011}
\begin{equation}
\begin{aligned}
 \overline{ \left\langle \sum_{j=1}^{n_1}  x^2 s_{jj}\right\rangle} =&
\frac{4k_x^2 \langle B \rangle^2 \langle T_1 \rangle\left((4-\beta)+\beta CV^2_{T_1}\right)}{3k_1}\\&+\frac{ 4k_x^2 \langle B \rangle^2 \langle T_2\rangle \left((3-\beta )+(1-\beta) CV^2_{T_2}\right)}{3 k_1}+\frac{2 k_x\langle B \rangle(2-\beta)}{3k_1}.\label{all higher partition01}
\end{aligned} 
\end{equation}
By having this term, the moments  $ \overline{\langle x^2g_{ij} \rangle}$ are derived by solving equations \eqref{higher_moment_partition1} and \eqref{higher_moment_partition2} in steady-state
\begin{equation}
\begin{aligned}
 \overline{  \langle x^2g_{ij} \rangle} = & \frac{4 k_x^2 \langle B \rangle^2 \langle\langle T_1 \rangle\left((4-\beta)+\beta CV^2_{T_1}\right)}{3k_2}p_i+\frac{ 4 k_x^2 \langle B \rangle^2 \langle T_2\rangle \left((3-\beta)+(1-\beta) CV^2_{T_2}\right)}{3 k_2}p_i\\
&
+\frac{8k_x^2 \langle B \rangle^2}{k_2}p_i\left( \frac{(1-\beta)j^2 +j}{i} \right)
+\frac{2 \alpha k_x\langle B \rangle(2-\beta)}{3k_2} p_i, \\& i= \{1,\ldots,n_2\},   \ j= \{1,\ldots,i\}.
\end{aligned} 
\end{equation}
Note that 
\begin{equation}
\overline{  \langle x^2 \rangle}= \left \langle x^2 \left(\sum_{i=1}^{n_1}\sum_{j=1}^{i}  s_{ij}+ \sum_{i=1}^{n_2}\sum_{j=1}^{i} g_{ij} \right)\right \rangle =\sum_{i=1}^{n_1} \sum_{j=1}^{i} \overline{  \langle  x^2 s_{ij} \rangle}+\sum_{i=1}^{n_2} \sum_{j=1}^{i} \overline{  \langle  x^2 g_{ij} \rangle} ,
\end{equation}
hence the second-order moment is
\begin{equation}
\begin{aligned}
\overline{  \langle x^2 \rangle}= & \frac{4 \langle T_1^3\rangle+16 \langle T_2^3 \rangle +2 \langle T\rangle^3 (3(2-\beta)^2+ \beta^2(5-3 \beta) CV^2_{T_1} +8(1-\beta)^2  CV^2_{T_2})}{3 \langle T\rangle}\\& +\frac{2 \alpha  k_x\langle B \rangle(2-\beta)  \langle T \rangle}{3}.
\end{aligned}
\end{equation}
Coefficient of variation squared gives noise raised from partitioning and cell-cycle variations, which subtracting equation \eqref{cv_division_time} from results gives partitioning noise as
\begin{align}
&CV^2_R =  \frac{4\alpha (2-\beta)}{3\left((\beta^2 -4 \beta +6)  + \beta^2CV^2_{T_1}+ 2(1-\beta)^2 CV^2_{T_2} \right)}\frac{1}{\overline{ \langle x \rangle}}.\label{cv_partitioning01}
\end{align}

\newpage
\subsection{Noise in protein count level contributed from stochastic production}
In order to calculate the noise caused by stochastic birth of protein, we use the model introduced in Section C.1. For this model, moments dynamics of $\langle x^2 \rangle$, $\langle x^2 s_{ij} \rangle$, and $\langle x^2 g_{ij} \rangle$ can be written as
\begin{small}
\begin{subequations}
\begin{align}
&\frac{d\langle x^2 \rangle}{dt}=
k_x \langle B^2 \rangle(2-\beta) +2 k_x \langle B \rangle \left( \langle x \rangle+ \left \langle  \sum_{i=1}^{n_2}  \sum_{j=1}^i x g_{ij}   \right \rangle\right)
+\frac{1}{4}\alpha k_2\left \langle\sum_{j=1}^{n_2} x  g_{jj} \right \rangle-\frac{3 k_2}{4}  \left \langle\sum_{j=1}^{n_2} x^2  g_{jj}\right  \rangle, \label{x2_birth}
\\
&\frac{d\langle x^2 s_{i1} \rangle}{dt}= \frac{k_x\langle B^2 \rangle \beta p'_i}{i}+  2 k_x \langle B \rangle \langle x s_{i1} \rangle + \frac{k_2}{4}  p'_i\left \langle \sum_{j=1}^{n_2}  x^2 g_{jj} \right \rangle + \frac{1}{4}\alpha k_2  p'_i\left \langle \sum_{j=1}^{n_2}  x g_{jj}\right  \rangle - k_1\langle x^2s_{i1} \rangle, \label{x2s1_birth}\\
&\frac{d\langle x^2 s_{ij} \rangle}{dt}=\frac{k_x\langle B^2 \rangle \beta p'_i}{i}+2 k_x \langle B \rangle \langle x s_{ij} \rangle 
- k_1\langle x^2 s_{ij} \rangle + k_1\langle x^2 s_{(i-1)j} \rangle, \ j= \left\lbrace 2,\ldots,i \right\rbrace,
\label{x2si_birth}\\
&\frac{d\langle x^2 g_{i1} \rangle}{dt}= \frac{2k_x\langle B^2 \rangle (1-\beta) p_i}{i}+ 4 k_x \langle B \rangle \langle x g_{i1} \rangle  + k_1 p_i\left \langle \sum_{j=1}^{n_1}  x^2 s_{jj} \right \rangle  - k_2\langle x^2g_{i1} \rangle, ,\label{x2g1_birth}\\
&\frac{d\langle x^2 g_{ij} \rangle}{dt}=\frac{2k_x\langle B^2 \rangle (1-\beta) p_i}{i}+ 4 k_x \langle B \rangle \langle x g_{ij} \rangle 
- k_2 \langle x^2 g_{ij} \rangle + k_2 \langle x^2 g_{(i-1)j} \rangle, \ j= \left\lbrace 2,\ldots,i \right\rbrace.
\label{x2gi_birth}
\end{align} 
\end{subequations}
\end{small}
The same as before we start by expressing the term $\overline{\langle \sum_{j=1}^{n_2}  x^2 g_{jj} \rangle}$, this term is calculated by analyzing equation \eqref{x2_birth} in steady-state
\begin{equation}
\begin{aligned}
 \overline{ \left  \langle\sum_{j=1}^{n_2}  x^2 g_{jj} \right  \rangle}  =& \frac{4k_x^2 \langle B \rangle^2 \langle T_1 \rangle\left((4-\beta)+\beta  CV^2_{T_1}\right)}{3k_2} +\frac{2 \alpha k_x \langle B \rangle (2-\beta)}{3 k_2}\\
&+\frac{16 k_x^2 \langle B \rangle^2 \langle T_2\rangle \left((3-\beta)+(1-\beta) CV^2_{T_1}\right)}{3 k_2}
+\frac{4 k_x(2-\beta)\langle B^2 \rangle}{3 k_2}.
\label{last gene birth}
\end{aligned}
\end{equation}
Substituting this term in equations \eqref{x2s1_birth} and \eqref{x2si_birth} results in 
\begin{equation}
\begin{aligned}
\overline{ \langle x^2 s_{ij} \rangle}=
 &\frac{k_x^2 \langle B \rangle^2 \langle T_1 \rangle\left((4-\beta )+\beta CV^2_{T_1}\right)}{3k_1}p'_i+\frac{ 4k_x^2 \langle B \rangle^2 \langle T_2\rangle \left((3-\beta)+(1-\beta) CV^2_{T_2}\right)}{3 k_1}p'_i\\
&+\frac{2k_x^2 \langle B \rangle^2}{k_1}p'_i\left( \frac{ \beta j^2 +(2-\beta)j}{i} \right)+\frac{2 \alpha k_x(2-\beta)\langle B \rangle}{3k_1}p'_i\\
& +\frac{ k_x \langle B^2 \rangle }{k_1} \left(\frac{2-\beta}{3} +\beta\frac{j}{i} \right)p'_i , \  i= \{1,\ldots,n_1\},  \ j= \{1,\ldots,i\}.\label{all x2si pbirth}
\end{aligned} 
\end{equation}
Similar to previous section, solving equations \eqref{x2g1_birth} and \eqref{x2gi_birth} gives the $\overline{\langle x^2g_{ij}\rangle}$ 
\begin{equation}
\begin{aligned}
\overline{ \langle x^2 g_{ij} \rangle }=&\frac{4 k_x^2 \langle B \rangle^2 \langle\langle T_1 \rangle\left((4-\beta )+\beta  CV^2_{T_1}\right)}{3k_2}p_i+\frac{ 4 k_x^2 \langle B \rangle^2 \langle T_2\rangle \left((3-\beta)+(1-\beta) CV^2_{T_2}\right)}{3 k_2}p_i\\
&
+\frac{8k_x^2 \langle B \rangle^2}{k_2}p_i\left( \frac{(1-\beta)j^2 +j}{i} \right)
+\frac{2 \alpha k_x\langle B \rangle(2-\beta)}{3k_2} p_i\\
& + \frac{ 2k_x\langle B^2 \rangle }{ k_2}\left( \frac{ 1+\beta }{3}  + (1-\beta)\frac{j}{i} \right)  p_i , \ i= \{1,\ldots,n_2\} ,  \ j= \{1,\ldots,i\}.\label{all higher birth}
\end{aligned} 
\end{equation}
Finally summing all the moments $\overline{ \langle x^2 s_{ij} \rangle}$, and $\overline{ \langle x^2 g_{ij} \rangle}$ results in $\overline{ \langle x^2\rangle}$ as
\begin{equation}
\begin{aligned}
\overline{ \langle x^2 \rangle }=&\frac{4 \langle T_1^3\rangle+16 \langle T_2^3 \rangle+2 \langle T\rangle^3 (3 a (-(\beta+1) \beta  CV^2_{T_1}+a-4)+8  CV^2_{T}+12)}{3 \langle T\rangle}\\& +\frac{2 \alpha  k_x\langle B \rangle(2-a)  \langle T \rangle}{3}+
k_x \langle B^2 \rangle\left(\frac{2-\beta}{3}  + \beta \left( \frac{1+CV_{T_1}^2}{2} \right) \right) \langle T_1 \rangle\\
&+2k_x \langle B^2 \rangle\left(\frac{1+\beta}{3}  + (1-\beta)\left( \frac{1+CV_{T_2}^2}{2} \right) \right) \langle T_2 \rangle.
\end{aligned}
\end{equation}
Steady-state analysis gives the noise from stochastic birth, random partitioning, and cell-cycle time variations. Subtracting noise of cell-cycle time and partitioning in equations \eqref{cv_division_time} and \eqref{cv_partitioning01} results in noise caused by stochastic production of protein 
\begin{align}
&CV^2_P =  \frac{(10-8\beta+3\beta^2) + 6(1-\beta)^2 CV^2_{T_2} + 3 \beta^2 CV^2_{T_1} }{3\left((\beta^2 -4 \beta +6)  +\beta^2CV^2_{T_1} + 2 (1-\beta)^2 CV^2_{T_2} \right)}\frac{\langle B^2 \rangle}{\langle B\rangle}\frac{1}{\overline{ \langle x \rangle}} .\label{cv_partitioning}
\end{align}

\newpage
\subsection{Effect of gene-duplication time in intrinsic noise}

\begin{figure}[!b]
\centering
\includegraphics[width=1\textwidth ,clip]{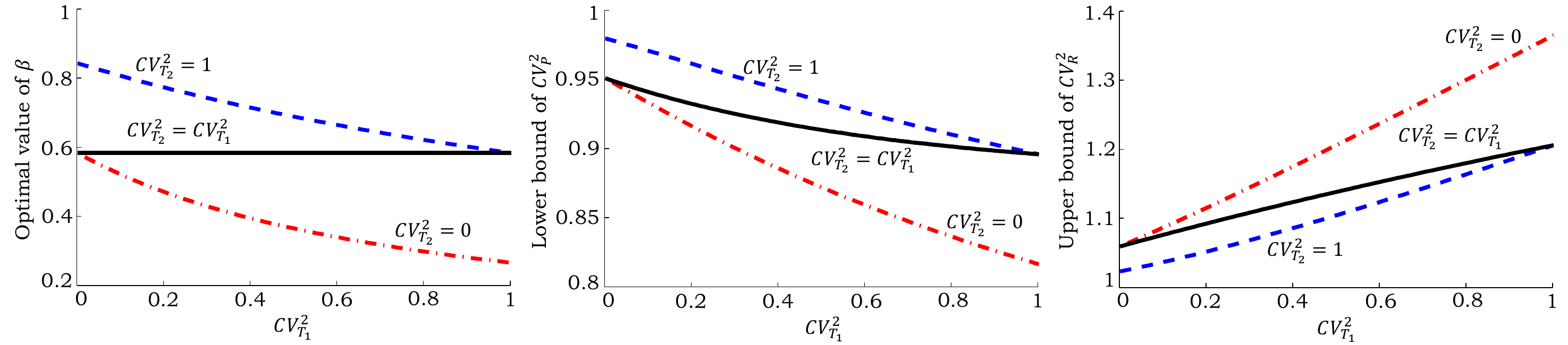} 
\caption*{Figure S3:
{\bf Effect of gene-duplication on intrinsic noise level}. \textit{Left}: Value of $\beta$ where $CV^2_P$ is minimized and $CV^2_R$ is maximized as a function of $CV^2_{T_1}$.
When $CV^2_{T_1}=CV^2_{T_2}$, noise levels always reach their extrema at $\beta=2-\sqrt{2}$. \textit{Middle \& Right}: Extremum values of $CV^2_P$ and $CV^2_R$ 
as a functions of $CV^2_{T_1}$. Noise levels are normalized by their values at $\beta=0$.
}
\label{fig8}
\end{figure}  

We investigate how the noise contributions from random partitioning and stochastic expression ($CV^2_{R}$ and $CV^2_{P}$ terms in equation (34) of the main text) change as 
$\beta$ is varied between $0$ and $1$. Results show that $CV^2_{R}$ and $CV^2_{P}$ follow the same qualitative shapes as reported in Fig. 6. There exists a $\beta^*$
\begin{footnotesize}
\begin{equation}
\beta^*= \frac{- \sqrt{2(2 CV^4_{T_1}+5 CV^2_{T_1} CV^2_{T_2}+3CV^2_{T_1} +2 CV^4_{T_2}+3 CV^2_{T_2} +1)}+2 CV^2_{T_1} +4 CV^2_{T_2} +2}{CV^2_{T_1} +2 CV^2_{T_2} +1},
\end{equation}
\end{footnotesize}
such that $CV^2_P$ is minimized and $CV^2_R$ is maximized when $\beta=\beta^*$. Note that when $CV^2_{T_1}=CV^2_{T_2}=0$, $\beta^*=2-\sqrt{2}$ as reported in the main text.
The minimum value of $CV^2_P$ and the maximum value of $CV^2_R$ are given by 
\begin{footnotesize}
\begin{equation}
\begin{aligned}
&CV^2_P= \frac{CV^2_{T_1} (3 CV^2_{T_2}+7)- \sqrt{2(2 CV^2_{T_1}+CV^2_{T_2}+1) (CV^2_{T_1}+2CV^2_{T_2} +1)}+7 CV^2_{T_2} +3}{3(CV^2_{T_1} (CV^2_{T_2}+3)+3 CV^2_{T_2}+1)}\frac{\langle B^2 \rangle}{\langle B\rangle}\frac{1}{\overline{ \langle x \rangle}},
\end{aligned}
\end{equation}
\end{footnotesize}
\begin{footnotesize}
\begin{equation}
\begin{aligned}
&  CV^2_R =\frac{\sqrt{2} \alpha  }{3\sqrt{(2 CV^2_{T_1}+CV^2_{T_2}+1) (CV^2_{T_1}+2 CV^2_{T_2}+1)}-3  \sqrt{2}  CV^2_{T_1} -3  \sqrt{2} CV^2_{T_2}},
\end{aligned}
\end{equation}
\end{footnotesize}
respectively.  Plots of  $\beta^*$ and optimal value of $CV^2_{R}$ and $CV^2_{P}$ as a function of $CV^2_{T_1}$ are shown in Fig. S4. Note that if noise in $T_1$ is high and $T_2$
is deterministic then $\beta^*$ shifts towards zero. Similarly, if noise in $T_2$ is high and $T_1$
is deterministic then $\beta^*$ shifts towards one.

\section{Noise level in unstable protein}
Consider an unstable protein with sufficiently high degradation rate $\gamma_x$ such that the protein level reaches steady-state instantaneously compared to the cell-cycle time (Fig. S4).
Let $\tau$ denote the time from the last division event, then
\begin{equation}
\begin{aligned}
&\overline{\langle x \vert \tau<T_1 \rangle }=  \frac{k_x \langle B \rangle}{\gamma_x} ,\\
&\overline{ \langle x \vert \tau>T_1 \rangle } =   \frac{2k_x \langle B \rangle}{\gamma_x},
\end{aligned} 
\end{equation}
where $T_1$ is the time in which duplication happens. The mean level of an unstable protein can be calculated as 
\begin{equation}
\overline{\langle x  \rangle}= \overline{\langle x \vert \tau < T_1 \rangle} p(\tau < T_1 ) +  \overline{\langle x \vert \tau > T_1 \rangle} p(\tau > T_1 ) ,
\end{equation}
where $ p(\tau < T_1 )$ and $ p(\tau > T_1 )$ denote the probability of being in the time interval before and after gene-duplication. Using 
\begin{equation}
p(\tau < T_1 ) = \beta, \ \  p(\tau > T_1 )=(1-\beta),
\end{equation}
we obtain
\begin{equation}
\overline{\langle x \rangle } = \frac{k_x \langle B \rangle (2-\beta)}{\gamma_x}. \label{mean}
\end{equation}

To compute the extrinsic noise component we consider deterministic protein production and decay. The second-order moment of $x(t)$ is given by
\begin{equation}
\begin{array}{l}
\overline{\langle x^2 \vert \tau<T_1 \rangle} =  \left(\frac{k_x \langle B \rangle}{\gamma_x}\right)^2 \\
\overline{\langle x^2 \vert \tau>T_1 \rangle} =  \left( \frac{2k_x \langle B \rangle}{\gamma_x}\right)^2\\
\end{array} \Rightarrow \overline{\langle x^2 \rangle } =\left( \frac{k_x \langle B \rangle }{\gamma_x}\right)^2 \beta +\left( \frac{2k_x \langle B \rangle }{\gamma_x}\right)^2 (1-\beta).
\label{x2 extrinsic}
\end{equation}
By using definition of $CV^2$, extrinsic noise is
\begin{equation}
CV^2_E= \frac{(1-\beta)\beta}{(2-\beta)^2}, \label{extrinsic}
\end{equation} 
which is zero at $\beta = 0,1$ and reaches its maximum at $\beta = 2/3$ (Fig. S4).

Next we compute the intrinsic noise component. If the protein decay is sufficiently high, the noise contribution from partitioning errors will be 
negligible because any errors will be instantaneously corrected due to rapid protein turnover. Noise raised from stochastic gene expression can be investigated by considering a model containing stochastic bursty production and stochastic degradation of proteins, where after gene-duplication the burst frequency doubles. Again assuming large enough $\gamma_x$,
 $\overline{\langle x^2 \vert \tau < T_1 \rangle} $ is equal to the steady-state second-order moment of a stochastic model with burst frequency $k_x$ (analyzed in \cite{moh01})
\begin{equation}
\begin{aligned}
&\overline{\langle x^2 \vert \tau < T_1 \rangle} = \left(\frac{k_x \langle B \rangle}{\gamma_x}\right)^2 + \frac{k_x \langle B^2 \rangle}{2 \gamma_x^2}  + \frac{k_x \langle B \rangle}{2 \gamma_x}.
\end{aligned}
\end{equation}
In comparison with equation \eqref{x2 extrinsic}, there are two extra terms at the right hand side of $\overline{\langle x^2 \vert \tau < T_1 \rangle}$. The first extra term is due to production of protein in random bursts and the second one is due to stochastic degradation of protein molecules.
Further for the same reasons (large degradation rate and rapid equilibration of the distribution), $\overline{\langle x^2 \vert \tau > T_1 \rangle} $ is equal to the second-order moment of a model containing stochastic bursty production of proteins with burst frequency $2 k_x$ which is
\begin{equation}
\begin{aligned}
&\overline{\langle x^2 \vert \tau > T_1 \rangle}= \left(\frac{2k_x \langle B \rangle}{\gamma_x}\right)^2 + \frac{k_x \langle B^2 \rangle}{ \gamma_x^2}  + \frac{k_x \langle B \rangle}{ \gamma_x}.
\end{aligned}
\end{equation}
Thus the second order moment of an unstable protein can be written as
\begin{equation}
\begin{aligned}
\overline{\langle x^2 \rangle } =&\left( \frac{k_x \langle B \rangle }{\gamma_x}\right)^2 \beta + \frac{k_x \langle B^2 \rangle }{2\gamma_x^2} \beta+ \frac{k_x \langle B\rangle }{2\gamma_x} \beta
\\&+\left( \frac{2k_x \langle B \rangle }{\gamma_x}\right)^2 (1-\beta)+\frac{k_x \langle B^2 \rangle }{\gamma_x^2} (1-\beta)+ \frac{k_x \langle B\rangle }{\gamma_x}(1-\beta).
\end{aligned}
\end{equation}
Using definition of $CV^2$ and subtracting extrinsic noise we obtain the following noise contribution from stochastic expression and decay \begin{equation}
CV^2_P=\frac{1}{2}\left( \frac{\langle B^2 \rangle}{\langle B\rangle}+1\right)\frac{1}{\overline{ \langle x \rangle}}. 
\label{production}
\end{equation}


\begin{figure}[!h]
\center
\includegraphics[width=1\textwidth]{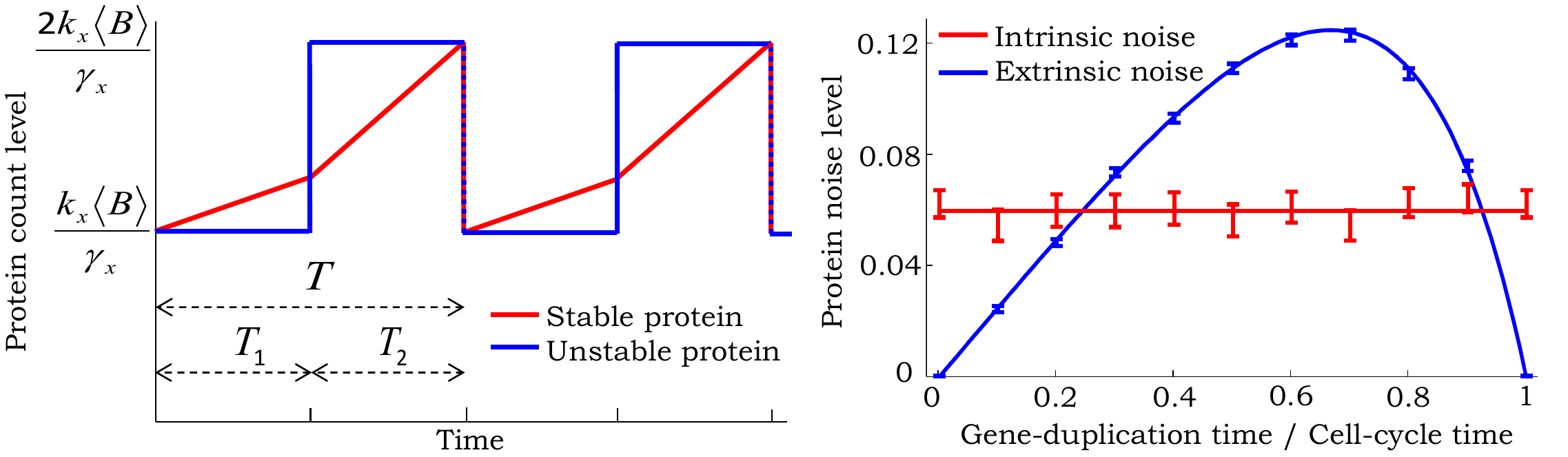}
\caption*{Figure S4: {\bf Contribution of gene duplication to noise levels of an unstable protein.} \textit{left}: For a stable protein, copy numbers accumulate in a bilinear fashion. In contrast, an unstable protein reaches equilibrium rapidly and its level changes in steps. \textit{Right}: Extrinsic and intrinsic noise predicted for an unstable protein as a function of $\beta$. Solid lines are predictions from \eqref{extrinsic} and \eqref{production}, which agree with estimates from $20,000$ Monte Carlo simulations. Parameters taken as 
 $\gamma_x=10 hr^{-1}$, and a geometric burst with $\langle B \rangle =6$. Burst frequency is changed to have a constant mean protein level of $100$ molecules for different values of $\beta$. $95\% $ confidence intervals are calculated via bootstrapping.
}
\label{fig:target distribution}
\end{figure}

\section*{Acknowledgements}

AS is supported by the National Science Foundation Grant DMS-1312926.

\bibliographystyle{plos2009}
\bibliography{thesisp,thesis,ref,thesismm}


\end{document}